%% file: pci.tex
\documentclass[journal]{new-aiaa}
\usepackage[utf8]{inputenc}
\usepackage{graphicx}
\usepackage{amsmath}
\usepackage{booktabs}
\usepackage{longtable,tabularx}
\usepackage{xcolor}
\usepackage{tikz}
\usetikzlibrary{patterns,calc,decorations.pathreplacing,decorations.pathmorphing,arrows.meta,matrix,positioning}
\usepackage[hidelinks]{hyperref}
\usepackage[capitalise]{cleveref}
\crefname{figure}{Fig.}{Figs.}
\Crefname{figure}{Figure}{Figures}
\crefname{equation}{Eq.}{Eqs.}
\Crefname{equation}{Equation}{Equations}
\crefname{section}{Sec.}{Secs.}
\Crefname{section}{Section}{Sections}
\crefname{appendix}{Appendix}{Appendices}

\graphicspath{{figures/}}

\newcommand{\ii}{\mathrm{i}}
\newcommand{\ee}{\mathrm{e}}
\newcommand{\dt}{\Delta t}
\newcommand{\herm}{\mathsf{H}}
\newcommand{\abs}[1]{\lvert #1\rvert}
\newcommand{\muh}{\hat{\mu}}
\newcommand{\betah}{\hat{\beta}}   %

\newcommand{\mat}[1]{\mathbf{#1}}     %
\newcommand{\gmat}[1]{\boldsymbol{#1}}  %
\newcommand{\snap}{\nu}  %
\newcommand{\chstar}{CH$^*$}
\newcommand{\oneW}{$1W$}
\newcommand{\twoW}{$2W$}

\title{Ensemble Floquet Decomposition of a Transverse Combustion Instability in a Swirl-Stabilized Flame}

\author{J.~Matt Quinlan\footnote{Assistant Professor, Department of Mechanical and Aerospace Engineering, 1420 Austin Bluffs Parkway. Member AIAA.}}
\affil{University of Colorado Colorado Springs, Colorado Springs, Colorado 80918}

\begin{document}

{\let\thefootnote\relax\footnotetext{Presented as Paper 2022-1857 at the AIAA SciTech 2022 Forum, San Diego, CA, and Virtual, January 3--7, 2022.}}
\maketitle

\begin{abstract}
Simultaneous 10\,kHz measurements of velocity, OH fluorescence, \chstar\ chemiluminescence, and acoustic pressure in a swirl-stabilized combustor are analyzed with the ensemble Floquet decomposition, a dynamic mode decomposition that advances one-period snapshot stacks by one period.
It identifies the 170\,Hz combustion instability, missed by dynamic mode decomposition, as the duct's first transverse mode, with the acoustically stiff injector at its velocity antinode.
In the proposed mechanism, the transverse acoustic velocity displaces the flame-anchoring stagnation point by about 0.4 nozzle radii, the swirling jet's shear layer carries the disturbance, likely amplified by its 1\,kHz mode, and the heat release responds over the whole flame, mainly on one side, near the acoustic pressure maximum.
The exhaust, near the pressure node, removes 30 to 100 times less energy from this mode than from its neighbors.
The 1\,kHz mode, a pair of helical vortices rotating through the fixed acoustic mode, is phase-coupled to it.
Only methods that see a whole rotation separate the instability, and of hundreds of duct configurations only this one was unstable.
We suggest that it is a parametric, acoustically coupled combustion instability, in which the rotating hydrodynamic mode periodically varies the coefficients of the acoustic mode.
\end{abstract}

\input{sections/00_nomenclature}
\input{sections/01_introduction}
\input{sections/02_facility}
\input{sections/03_methods}
\input{sections/04_results}
\input{sections/05_discussion}

\input{sections/06_conclusions}

\section*{Funding Sources}
This work was supported by the Air Force Office of Scientific Research under award FA9550-14-1-0051 (program officer Dr.\ Mitat Birkan).

\section*{Acknowledgments}
The author thanks Professor Ben T.\ Zinn, who advised the dissertation in which this work began, and Professor Luca Dieci, who pointed the way to the theory of linear periodic systems.
This work used the INCLINE cluster at the University of Colorado Colorado Springs; INCLINE is supported by the National Science Foundation, grant \#2017917.
A large language model (Claude, Anthropic) was used to edit the text and to help prepare some of the figures.
The author assumes responsibility for all content.

\bibliography{references}

\end{document}

%% file: sections/00_nomenclature.tex
\section*{Nomenclature}

{\renewcommand\arraystretch{1.0}
\noindent\begin{longtable*}{@{}l @{\quad=\quad} l@{}}
$\mat{A}$ & constant matrix that advances a snapshot by one sample \\
$A_d$, $A_j$ & duct cross-sectional area; effective area of the exhaust jets, m$^2$ \\
$C_d$ & discharge coefficient of the exhaust holes \\
$d_h$ & diameter of the exhaust holes, m \\
$c$ & speed of sound, m/s; as an index, phase within the analysis period, $1,\dots,p$ \\
$d$, $s$ & number of snapshots stacked in a column of a data matrix; shift between columns, samples \\
$E$ & period-averaged acoustic energy of a duct mode, J \\
$f$ & frequency, Hz \\
$H_{qp}$ & transfer function from the combustor dynamic pressure near the injector to the heat-release signal \\
$K$ & number of columns of the FD and EFD data matrices, the number of whole analysis periods in the record \\
$L_1$, $L_2$ & lengths of the duct on either side of the injector, m \\
$\mat{M}_c$ & monodromy matrix, which advances a snapshot at phase $c$ by one period \\
$\dot m$ & total mass flow rate, kg/s \\
$N$ & number of snapshots \\
$\widebar{P}$ & mean chamber pressure, Pa \\
$p$ & number of samples in the analysis period \\
$p'_1$, $p'_2$, $p'_3$ & dynamic pressure signals of the duct sensor and of the two injector sensors, Pa \\
$p'_i$ & combustor dynamic pressure near the injector, $(p'_2+p'_3)/2$, Pa \\
$q'$ & heat-release fluctuation, taken proportional to the \chstar\ signal \\
$r$ & truncation rank \\
$T$ & analysis period, s \\
$u$, $v$ & transverse and axial velocity components, m/s \\
$\mat{V}_{a:b}$ & matrix whose columns are the snapshots $a$ to $b$; $\mat{V}^{(c)}$, period-sampled at phase $c$; $\widetilde{\mat{V}}$, stacked \\
$W$ & period-averaged acoustic power removed by the exhaust, W \\
$x$, $y$ & transverse (duct) and axial (injector) coordinates, m \\
$\alpha$ & amplitude decay rate of a duct mode, s$^{-1}$ \\
$\betah$ & estimated eigenvalue of Hankel-DMD \\
$\gamma$ & ratio of specific heats \\
$\gamma^2$ & magnitude-squared coherence \\
$\dt$ & sampling interval, s \\
$\muh$ & estimated Floquet multiplier \\
$\snap_k$, $\tilde\snap_k$ & composite snapshot at time $k\,\dt$; stack of the $p$ snapshots of period $k$ \\
$\xi$ & acoustic displacement amplitude, m \\
$\rho$ & density, kg/m$^3$ \\
$\gmat{\Phi}_k$ & matrix that advances the state from sample $k$ to sample $k+1$ \\
$\tilde\psi_l$, $\hat\psi_{l,j}$ & EFD mode $l$; its $j$th harmonic over the analysis period \\
$\omega$ & angular frequency, rad/s \\
$nW$ & $n$th acoustic mode along the transverse duct \\
\end{longtable*}}

%% file: sections/01_introduction.tex
\section{Introduction}

\lettrine{C}{ombustion} instabilities arise when unsteady heat release adds energy to an acoustic mode of the combustor faster than the mode loses it \cite{Rayleigh-1878,Culick-2006,Lieuwen-2005}.
They have affected most rocket engine development programs \cite{Harrje-1972,Yang-1995}.
Their transverse forms, in which the acoustic velocity is perpendicular to the injected flow, are the most damaging in rocket engines \cite{Harrje-1972} and have been studied less than the longitudinal forms in swirl-stabilized combustors \cite{OConnor-2015}.
There is an apparent paradox in how they are sustained.
Even at damaging amplitudes the acoustic pressure and velocity are small compared with their mean values almost everywhere, which makes a direct forcing of the combustion unlikely.
What is needed is a hydrodynamic instability that turns the small acoustic perturbation into a large motion of the flow and the flame \cite{Poinsot-1987,Schadow-1992,Ducruix-2003,Candel-2014}.
The acoustic velocity is significant only where the flow is slow, at stagnation points and on the low-speed sides of shear layers, and it is there that the hydrodynamics can be excited \cite{Quinlan-2022}.
Establishing how this happens in a given combustor requires the velocity field, the flame, and the acoustics to be observed together and resolved in time.

The facility studied here was built for that purpose \cite{Quinlan-2017,Quinlan-2015}.
A single swirl injector fires across a duct closed at both ends by movable pistons.
The positions of the pistons set the length of the duct on either side of the injector, and with it the frequencies of the transverse modes and the place of the injector in each mode, anywhere between a pressure antinode and a velocity antinode.
The duct has a nearly constant cross-section, so that its transverse acoustic field is nearly one-dimensional below the cutoff frequency of the first acoustic mode across the duct, and its lowest modes lie at hundreds of hertz, as do the transverse instabilities of full-scale liquid rocket engines, rather than at the several kilohertz of sub-scale rocket combustors \cite{Quinlan-2015}.
The injector is acoustically stiff, so that its flow does not respond to the acoustic pressure in the chamber and the injection-coupled route to instability is closed, and the exhaust is directly above it, so that a mode with its velocity antinode at the injector has its pressure node at the exhaust and loses little energy there.
Velocity, OH fluorescence, \chstar\ chemiluminescence, and acoustic pressure were recorded simultaneously at $10$\,kHz.
For one duct length the combustion in the rig is unstable, with a narrow, high peak in the dynamic pressure at $170$\,Hz; for a shorter duct, at nearly the same operating conditions, it is stable.
The dynamic mode decomposition (DMD) of the combined measurements nevertheless returned no mode at $170$\,Hz \cite{Quinlan-2015,Quinlan-2017}.
Its leading mode was instead near $1$\,kHz, a pair of helical vortices spinning around the inner recirculation zone of the swirling jet \cite{Quinlan-2017}.

DMD \cite{Schmid-2010} fits one constant linear map from each snapshot to the next, and in the paper that introduced it Schmid observed that snapshots sampled at the period of an oscillatory flow would instead identify the Floquet matrix of that flow \cite[\S 2.6]{Schmid-2010}.
A companion paper \cite{Quinlan-2026-FD} develops that observation into two methods and analyzes them: the Floquet decomposition (FD), which is DMD applied to the record subsampled once per period, and the ensemble Floquet decomposition (EFD), which stacks all the samples of one period into each snapshot and advances the stack by one period.
For a linear time-periodic system both are exact, EFD under the weaker condition that the measurements over one period, rather than at one instant, determine the state; the methods were developed in \cite{Quinlan-2015}, and a preliminary account of their application to the present data appeared in \cite{Quinlan-2022}.
EFD stacks snapshots as delay-embedded (Hankel) DMD does \cite{Arbabi-2017,LeClainche-2017}, but the two come from different views of DMD.
Hankel-DMD keeps the view of a time-invariant system: it fits one matrix that advances the data by one sample and represents a mode of a periodic system by several eigenvalues, one for each harmonic of the period present in the mode.
EFD takes the system to be periodic and fits the matrix that advances the data by one whole period, the monodromy matrix of Floquet theory, so that it returns each Floquet multiplier once, as the number by which the mode grows or decays in one period, with the mode resolved in the phase of the period.
With a window of one period the two contain the same information \cite{Quinlan-2026-FD}, but only EFD presents it in the form in which Floquet theory decides stability.

Floquet theory is the theory of linear systems with periodic coefficients, and so of parametric instability.
A parametric instability is one in which a periodic variation of the coefficients of a linear system, rather than a forcing term, makes a solution grow.
The standard example is the pendulum whose pivot oscillates vertically, which for small angles obeys Mathieu's equation \cite{Kovacic-2018,Verhulst-2011},
\begin{equation}
  \ddot\theta + \left[\kappa + \epsilon\cos(\omega_\mathrm{f}t)\right]\theta = 0 ,
  \label{eq:mathieu}
\end{equation}
and is unstable in tongues of the $(\kappa,\epsilon)$ plane that issue from $\omega_\mathrm{f} = 2\sqrt{\kappa}/n$, $n = 1,2,\dots$, the strongest at twice the natural frequency.
Stability is decided by the Floquet multipliers, the eigenvalues of the map that advances the state by one period \cite{Floquet-1883,Hale-1980}.
The connection with combustion instability is the hydrodynamic amplifier.
The shear layers and vortices of an injector flow oscillate at their own frequencies, and with them the velocity, the position of the flame, and the sound speed, which enter the equations of the acoustic field as coefficients.
To an acoustic mode these oscillations are what the moving pivot is to the pendulum, and parametric, acoustically coupled combustion instabilities are therefore possible \cite{Quinlan-2022}.

Parametric instability has a long history in combustion.
Markstein and Squire reduced the cellular response of flames vibrating in tubes, at half the acoustic frequency, to Mathieu's equation \cite{Markstein-1953,Markstein-1955,Markstein-1964}, Searby and Rochwerger measured its threshold in agreement with theory \cite{Searby-1991}, and later work treats it as the violent secondary stage of the acoustic instability of flames in tubes \cite{Pelce-1992,Clanet-1999,Bychkov-1999}; the same mechanism has been predicted for a shear layer in an acoustic field \cite{Bilgili-2020}.
In all of this work the acoustic field modulates the flame front or the shear layer, which responds subharmonically.
In swirl-stabilized flames, Terhaar et al.\ \cite{Terhaar-2016} modeled the precessing vortex core as an oscillator forced parametrically by the acoustic velocity, Steinberg et al.\ \cite{Steinberg-2010} and Moeck et al.\ \cite{Moeck-2012} observed heat release at the difference between the frequencies of a helical vortex and a thermoacoustic mode, Steinberg et al.\ \cite{Steinberg-2012} showed that the helical vortex can drive or damp the thermoacoustic mode, and Karmarkar et al.\ \cite{Karmarkar-2022} found that a precessing vortex core suppresses a thermoacoustic mode as their frequencies approach each other.

This paper applies EFD to the two configurations, with the period of the shear-layer mode near $1$\,kHz as the analysis period.
EFD identifies the instability at $170$\,Hz as one mode whose velocity, OH, \chstar, and dynamic pressure parts share one multiplier, which is what allows the fields to be related to the acoustics.
From it we develop the feedback loop proposed in outline in \cite{Quinlan-2022}: the first transverse mode of the duct acts on the flame through its transverse velocity at an acoustically stiff injector, the disturbance is carried and amplified along the shear layer of the swirling jet, and the heat release returns energy to the mode, which loses little through an exhaust at its pressure node.
The shear-layer mode is a pair of helical vortices rotating through the fixed acoustic mode, the two are coupled, and taken together the evidence suggests that the instability is a parametric, acoustically coupled combustion instability.

After the experiment and the methods (\cref{sec:experiment,sec:methods}), \cref{sec:results} presents the results, from the decomposition to the tests for a parametric coupling, and \cref{sec:discussion} assembles the mechanism and discusses whether it is parametric.
Relative to \cite{Quinlan-2022}, the analysis of the stable configuration and of a cold-flow record, a corrected time alignment of the dynamic pressure and optical records, the velocity data reprocessed to resolve the acoustic motion, the loss model, and the comparison with Hankel-DMD are new, and with them the evidence that the instability is sustained by a feedback through two widely separated frequencies, which is what points to a parametric, acoustically coupled combustion instability.

%% file: sections/02_facility.tex
\section{Experiment}
\label{sec:experiment}

\subsection{Combustor}
\label{sec:rig}

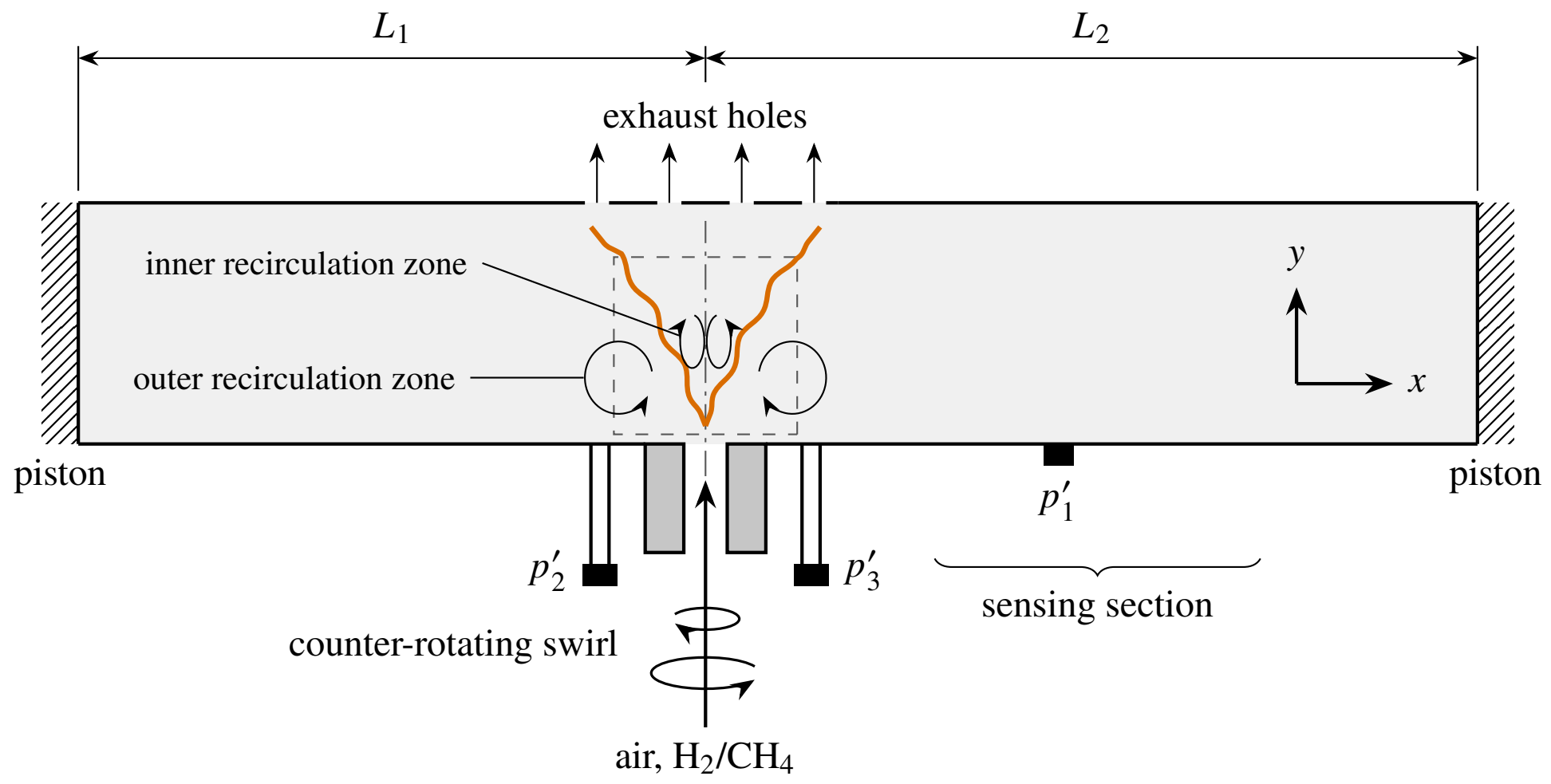
\begin{figure}[tb]
\centering
\begin{tikzpicture}[x=1cm,y=1cm,xscale=-1,font=\small,>=Stealth]
  \fill[gray!12] (-6.4,0) rectangle (5.2,2.0);
  \draw[thin,gray!70!black,dash pattern=on 6pt off 2pt on 1pt off 2pt] (0,-2.2) -- (0,1.85);
  \draw[thin,gray!70!black,dashed] (-0.76,0.08) rectangle (0.76,1.55);   %
  \draw[thick] (-6.4,0) -- (-0.5,0) (0.5,0) -- (5.2,0);
  \draw[thick] (-6.4,2.0) -- (-1.1,2.0) (1.1,2.0) -- (5.2,2.0);
  \fill[pattern=north east lines] (-6.7,0) rectangle (-6.4,2.0);
  \fill[pattern=north east lines] (5.2,0) rectangle (5.5,2.0);
  \draw[thick] (-6.4,0) -- (-6.4,2.0) (5.2,0) -- (5.2,2.0);
  \node[below] at (-6.55,0) {piston};
  \node[below] at (5.35,0) {piston};
  \draw[thick,fill=gray!45] (-0.5,-0.9) rectangle (-0.18,0);
  \draw[thick,fill=gray!45] (0.18,-0.9) rectangle (0.5,0);
  \draw[->,thick] (0,-2.35) -- (0,-0.3);
  \node[below] at (0,-2.35) {air, H$_2$/CH$_4$};
  \draw[->,semithick] (0,-1.45) ++(25:0.28 and 0.09) arc[start angle=25,end angle=335,x radius=0.28,y radius=0.09];
  \draw[->,semithick] (0,-1.9) ++(155:0.45 and 0.13) arc[start angle=155,end angle=-155,x radius=0.45,y radius=0.13];
  \node[left] at (0.6,-1.68) {counter-rotating swirl};
  \draw[very thick,orange!85!black,decorate,decoration={snake,amplitude=1pt,segment length=15pt,pre length=3pt,post length=1pt}] (0,0.15) .. controls (-0.25,0.8) and (-0.55,1.3) .. (-0.95,1.8);
  \draw[very thick,orange!85!black,decorate,decoration={snake,amplitude=1pt,segment length=15pt,pre length=3pt,post length=1pt}] (0,0.15) .. controls (0.25,0.8) and (0.55,1.3) .. (0.95,1.8);
  \draw[->,thin] (0.11,0.85) ++(100:0.1 and 0.22) arc[start angle=100,end angle=420,x radius=0.1,y radius=0.22];
  \draw[->,thin] (-0.11,0.85) ++(80:0.1 and 0.22) arc[start angle=80,end angle=-240,x radius=0.1,y radius=0.22];
  \draw[->,thin] (0.72,0.55) ++(170:0.28 and 0.3) arc[start angle=170,end angle=-150,x radius=0.28,y radius=0.3];
  \draw[->,thin] (-0.72,0.55) ++(10:0.28 and 0.3) arc[start angle=10,end angle=330,x radius=0.28,y radius=0.3];
  \draw[thin] (0.22,0.9) -- (1.85,1.5) node[left,font=\footnotesize] {inner recirculation zone};
  \draw[thin] (1.01,0.55) -- (1.95,0.55) node[left,font=\footnotesize] {outer recirculation zone};
  \foreach \xx in {-0.9,-0.3,0.3,0.9} {\draw[->] (\xx,2.0) -- (\xx,2.5);}
  \draw[thick] (-1.1,2.0) -- (-1.0,2.0) (-0.8,2.0) -- (-0.4,2.0) (-0.2,2.0) -- (0.2,2.0) (0.4,2.0) -- (0.8,2.0) (1.0,2.0) -- (1.1,2.0);
  \node[above] at (0,2.5) {exhaust holes};
  \draw[thick] (-0.95,0) -- (-0.95,-1.0) (-0.8,0) -- (-0.8,-1.0);
  \draw[thick] (0.8,0) -- (0.8,-1.0) (0.95,0) -- (0.95,-1.0);
  \fill (-1.02,-1.18) rectangle (-0.73,-1.0);
  \fill (0.73,-1.18) rectangle (1.02,-1.0);
  \node[right] at (-1.02,-1.05) {$p'_3$};
  \node[left] at (1.02,-1.05) {$p'_2$};
  \fill (-3.05,-0.18) rectangle (-2.8,0);
  \node[below] at (-2.92,-0.18) {$p'_1$};
  \draw[decorate,decoration={brace,amplitude=4pt}] (-4.6,-0.95) -- (-1.9,-0.95) node[midway,below=4pt] {sensing section};
  \draw[<->] (-6.4,3.2) -- (0,3.2) node[midway,above] {$L_2$};
  \draw[<->] (0,3.2) -- (5.2,3.2) node[midway,above] {$L_1$};
  \draw[thin] (-6.4,2.1) -- (-6.4,3.3) (5.2,2.1) -- (5.2,3.3) (0,3.0) -- (0,3.3);
  \draw[->,thick] (-4.9,0.5) -- (-5.7,0.5) node[right] {$x$};
  \draw[->,thick] (-4.9,0.5) -- (-4.9,1.3) node[above] {$y$};
\end{tikzpicture}
\caption{Schematic of the combustor and transverse duct, not to scale, drawn from the side of the OH and \chstar\ cameras, with the sensing section and positive $x$ on the right. The duct sensor $p'_1$ is at $x = 336$\,mm, and the injector sensors $p'_2$ and $p'_3$ are at $x = -26.3$ and $+26.3$\,mm at the closed ends of standoff tubes that open on the injector face. The looped arrows mark the inner and outer recirculation zones of the swirling jet and the opposite senses of rotation of the two air passages. The dash-dot line is the injector axis, and the dashed rectangle the field of the optical measurements.}
\label{fig:rig}
\end{figure}

The facility is described in detail in \cite{Quinlan-2017,Quinlan-2015}; \cref{fig:rig} shows the parts that matter here.
A single swirl injector in the lower wall of the duct injects transverse to it, the products leave through eight holes of $6.35$\,mm diameter in the upper wall, in two rows of four centered above the injector, and movable pistons close the duct at both ends, so that its acoustic modes of interest, the widthwise modes $1W$, $2W$, and so on \cite{Quinlan-2015}, are standing waves along the duct, transverse to the injector axis.
The combustor and the adjacent sensing section are $76.2$\,mm high and $59.9$\,mm deep and join round ducts of $73.7$\,mm internal diameter, so the cross-section is nearly constant along the whole duct; the combustor has windows on its front and back faces and in its upper wall for the laser sheets.

The origin of coordinates is at the center of the injector exit; $x$ runs along the duct, positive toward the sensing section, which is to the right in \cref{fig:rig}, and $y$ along the injector axis.
The duct has length $L_2 = 1.016$\,m on the side of the sensing section and an adjustable length $L_1$ on the other, and two configurations are analyzed, $L_1 = 0.262$\,m, called \emph{stable}, and $L_1 = 0.914$\,m, called \emph{unstable}, at otherwise nearly the same operating conditions (\cref{tab:conditions}).
Hundreds of other piston settings, with the \oneW\ and \twoW\ at many different frequencies and often with the injector near a velocity antinode, were all stable: the unstable configuration analyzed here is the only one that produced a narrow, tonal peak in the dynamic pressure.

The injector has two counter-rotating swirling air passages, with about one fifth of the air in the outer one, and a nozzle of $10.9$\,mm diameter.
The fuel is hydrogen doped with methane, the methane serving to produce \chstar\ chemiluminescence.
The flows of \cref{tab:conditions} correspond to global equivalence ratios of $1.00$ and $1.03$ in the stable and unstable configurations and, with the lower heating values of the two fuels, to thermal powers of $37$ and $38$\,kW.

\begin{table}[tb]
\caption{Operating conditions}
\label{tab:conditions}
\centering
\begin{tabular}{lccccccc}
\toprule
Configuration & $L_1$, m & $L_1+L_2$, m & $\dot m_\mathrm{air}$, g/s & $\dot m_{\mathrm{H}_2}$, g/s & $\dot m_{\mathrm{CH}_4}$, g/s & Outer swirl air, \% & $T_\mathrm{air}$, K \\
\midrule
Stable   & $0.262$ & $1.28$ & $11.56$ & $0.188$ & $0.297$ & $20.8$ & $327.4$ \\
Unstable & $0.914$ & $1.93$ & $11.58$ & $0.189$ & $0.313$ & $21.1$ & $316.6$ \\
\bottomrule
\end{tabular}
\end{table}

\subsection{Diagnostics}
\label{sec:diagnostics}

Dynamic pressure was measured with piezoelectric sensors (Kistler 211B5) and the manufacturer's amplifiers, low-pass filtered at $8$\,kHz by four-pole analog filters, and sampled at $20$\,kHz for $4$\,s.
Three signals are used (\cref{fig:rig}).
The duct sensor $p'_1$ is in the sensing section at $x = 336$\,mm.
The injector sensors $p'_2$ and $p'_3$ are on either side of the nozzle at $x = \mp 26.3$\,mm; to protect them from the flame they sit at the closed ends of standoff tubes, $6.35$\,mm in diameter and about $85$\,mm long, that open on the injector face.
These tubes are quarter-wave resonators, with consequences described in \cref{sec:spectra}.
The mean of the two injector signals, the combustor dynamic pressure near the injector, $p'_i = (p'_2+p'_3)/2$, is used as the acoustic pressure at the flame.
The amplifiers were set to a sensitivity of $1$\,psi/V.
The data acquisition recorded the trigger of the cameras with the dynamic pressure, and the dynamic pressure records are brought to the optical frame rate of $10$\,kHz by keeping every second sample, starting with the sample taken at the first trigger, so that each frame is paired with the dynamic pressure measured at the same instant; where a digital filter is applied (\cref{sec:snapshots}), it is applied without phase shift to the $20$\,kHz record before the samples are taken.
The time derivative of each signal is formed at the same instants by a five-point central difference of the $20$\,kHz record.
The acoustic pressure and its rate of change at one point give, at a single instant, both the amplitude and the phase of an oscillation there, as the position and velocity of an oscillator do, so the derivatives double what the three sensors contribute to each snapshot about the state of the acoustic field.

Particle image velocimetry (PIV) used a dual-cavity Nd:YLF laser and TiO$_2$ seed of $1$\,$\mu$m nominal diameter; image pairs $9$\,$\mu$s apart were recorded at $10$\,kHz and give the velocity components $u$ and $v$ in the plane through the injector axis that contains the duct axis, over $-29 \le x \le 29$\,mm and $0 \le y \le 59$\,mm, on a grid of $0.89$\,mm.
OH planar laser-induced fluorescence (PLIF) was excited near $283.95$\,nm in the same plane and recorded through an intensifier with a $160$\,ns gate.
\chstar\ chemiluminescence was collected through a $430$\,nm filter of $10$\,nm width and an intensifier with a $90$\,$\mu$s gate; it is integrated along the line of sight, through the depth of the duct.
The three cameras were driven by one master trigger at $10$\,kHz: in each cycle the PIV camera recorded a pair of frames and each of the other two cameras one frame, with the PLIF pulse placed between the two PIV pulses.
The first $35$ frames of each record, taken while the lasers settled, are discarded, which leaves $N = 8504$ simultaneous frames, $0.85$\,s.
In the earlier analyses of these data \cite{Quinlan-2015,Quinlan-2022} the dynamic pressure record was not shifted to match, so that each frame was paired with the dynamic pressure $3.5$\,ms earlier; here each frame is paired with the dynamic pressure sample of the same instant.
Further details of the equipment are in \cite{Quinlan-2017,Quinlan-2015}.

The PIV camera viewed the flow through one window and the other two cameras through the opposite one.
Every field in this paper is shown as the OH and \chstar\ cameras saw it, the velocity fields being mirrored to match, and \cref{fig:rig} is drawn from the same side: the sensing section and positive $x$ are to the right.

The velocity vectors of the unstable configuration were computed for this paper from the raw images, because those computed at the time of the experiment with commercial software \cite{Quinlan-2015,Quinlan-2022} contain outliers and are prone to peak locking.
The images were reprocessed by iterative window deformation, with the time-averaged field of the original vectors as the first estimate, which lets the windows correlate where a fast jet passes within a few millimeters of reverse flow near the nozzle; $85\%$ of the vectors are accepted, and rejected vectors are replaced by the median of their neighbors in the composite snapshots only.
A time-averaged first estimate pulls the instantaneous vectors slightly toward the mean, so the amplitudes of fluctuations in the jets may be underestimated by up to a fifth; their phases are not affected.
The spatial averages of \cref{sec:res-mode} use the whole image, about $95$ by $63$\,mm; the composite snapshots and the maps use the field given above.
For the stable configuration only the original vectors are used.

\subsection{Composite snapshots}
\label{sec:snapshots}

The decompositions act on composite snapshots, as in \cite{Richecoeur-2012}: the snapshot $\snap_k$ is the concatenation of the two velocity components, the OH image, the \chstar\ image, and the dynamic pressure data at one instant, so that each mode has velocity, scalar, and dynamic pressure parts that share one eigenvalue.
The dynamic pressure part consists of the three signals, band-pass filtered between $80$ and $500$\,Hz without phase shift (fourth-order Butterworth, applied forward and backward), and of their time derivatives, formed from the filtered signals.
The band keeps the low-frequency part of the dynamic pressure spectrum, in which the \oneW\ and \twoW\ lie, and excludes the standoff resonance near $1$\,kHz that dominates the spectra of the injector sensors.
Each of the four kinds of data is multiplied by a scale factor, chosen so that the norm of its time-averaged absolute value equals a prescribed number: $\frac13$, $1$, $1$, and $3$ for the velocity, OH, \chstar, and dynamic pressure data in the baseline.
The dynamic pressure is weighted most heavily because it alone was filtered before sampling and so is free of aliasing, and the velocity least because it is the noisiest.
The band-pass and the weights are choices of observable.
A linear filter applied to an observable moves no eigenvalue of a linear system, but it is a prior selection of frequencies, and \cref{sec:res-efd} reports how the results depend on it.

\subsection{Dynamic pressure spectra and the standoff resonance}
\label{sec:spectra}

\begin{figure}[tb]
\centering
\includegraphics[width=\textwidth]{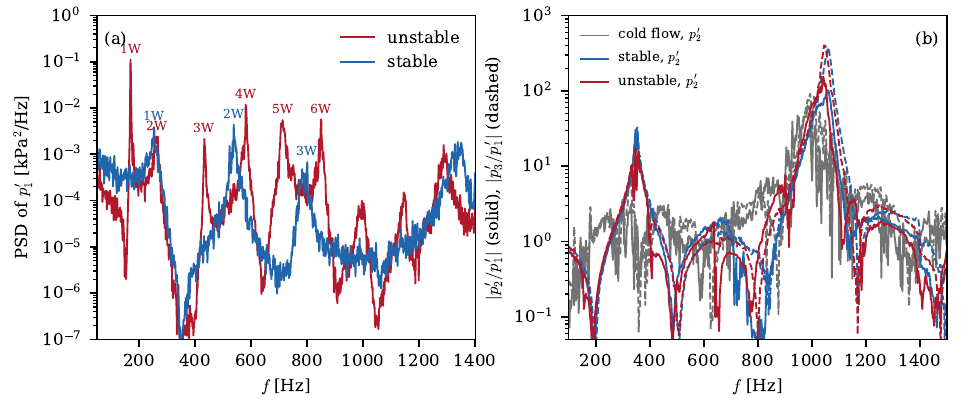}
\caption{Dynamic pressure spectra. a) Power spectral density of the duct sensor $p'_1$ in the two configurations, with the duct modes labeled. b) Magnitude of the transfer function from the duct sensor to each injector sensor, for the two configurations and for cold flow.}
\label{fig:spectra}
\end{figure}

\Cref{fig:spectra}(a) shows the spectrum of the duct sensor in the two configurations.
In the unstable configuration the modes of the duct appear as a ladder of peaks at $170$, $264$, $433$, $581$, $713$, and $850$\,Hz.
Above the second, their spacing, about $140$\,Hz, is that of a closed duct of length $L_1+L_2$ with an effective sound speed of $549$\,m/s \cite{Quinlan-2015}.
The first two are displaced from that ladder, the first upward from $142$ to $170$\,Hz, an effect of the nonuniform temperature along the duct that a one-dimensional model reproduces only in part \cite{Quinlan-2015}.
The $1W$ peak is the largest and by far the narrowest, $2.3$\,Hz at half power against $4$ to $14$\,Hz for the others, and the band within $10$\,Hz of it holds $45\%$ of the variance of the dynamic pressure at the duct sensor between $100$ and $1400$\,Hz.
In the stable configuration the duct is shorter, the ladder starts at $254$\,Hz, and no peak stands out.

The spectra of the injector sensors have these peaks and, in addition, a broad peak near $1$\,kHz that is their largest feature in both configurations.
It lies in the band of the mode of the injector shear layer examined in \cref{sec:res-khz}, but the standoff tubes modify what the sensors record there.
A tube closed at one end resonates at a quarter wavelength, and the ratio of each injector signal to the duct signal (\cref{fig:spectra}b) shows that resonance near $1050$\,Hz in the reacting configurations and near $990$\,Hz in a record taken in cold flow, with air only, which matches the length of the tubes at room temperature; the gas in the tubes is warmer during combustion, although the tubes are air-cooled to protect the sensors.
The peak near $1$\,kHz in the injector signals therefore locates the band of the activity at the injector, but not its frequency or width.
Well below the resonance the tubes change the amplitude only slightly and not its sign, so the injector signals are faithful at the first two duct modes; above about $500$\,Hz their amplitudes are not.

%% file: sections/03_methods.tex
\section{Methods of Analysis}
\label{sec:methods}

\subsection{Ensemble Floquet decomposition}
\label{sec:efd}

\subsubsection{Dynamic mode decomposition}
\label{sec:dmd}

The composite snapshots $\snap_k$ collect all the data measured at the time $k\,\dt$ into one column of $m$ entries (\cref{sec:snapshots}), and the record consists of $N$ of them.
Let $\mat{V}_{a:b}$ denote the matrix whose columns are the snapshots $a$ to $b$.
DMD \cite{Schmid-2010,Tu-2014} looks for one constant $m\times m$ matrix $\mat{A}$ that advances every snapshot to the next, $\snap_{k+1} = \mat{A}\snap_k$ for $k = 1,\dots,N-1$, which in terms of the data matrices is the single requirement
\begin{equation}
  \mat{V}_{2:N} = \mat{A}\,\mat{V}_{1:N-1} \quad\text{with}\quad \mat{V}_{a:b} = \begin{bmatrix} \snap_a & \snap_{a+1} & \cdots & \snap_b \end{bmatrix} .
  \label{eq:dmd}
\end{equation}
Such a matrix exists when the snapshots are the state of a linear system whose coefficients do not change with time.
If the state obeys $\dot\snap = \mat{L}\snap$ with a constant matrix $\mat{L}$, then $\mat{A} = \ee^{\mat{L}\dt}$, and each eigenvalue of $\mat{A}$ is $\lambda = \ee^{(\sigma+\ii\omega)\dt}$, where $\sigma+\ii\omega$ is an eigenvalue of $\mat{L}$, with growth rate $\sigma$ and angular frequency $\omega$.
The matrix $\mat{A}$ is not known, and with $m$ in the tens of thousands it is never formed.
DMD obtains its eigenvalues from the data in four steps.

First, the singular value decomposition (SVD) of the first data matrix, truncated to its $r$ largest singular values,
\begin{equation}
  \mat{V}_{1:N-1} \approx \mat{U}\gmat{\Sigma}\mat{W}^{\herm} , \qquad \mat{V}_{1:N-1}^{+} = \mat{W}\gmat{\Sigma}^{-1}\mat{U}^{\herm} ,
  \label{eq:svd}
\end{equation}
writes the snapshots in terms of $r$ orthonormal directions, the columns of the $m\times r$ matrix $\mat{U}$, which are the proper orthogonal modes of the data.
The diagonal matrix $\gmat{\Sigma}$ holds the singular values, which measure how much the data vary along each direction, and the columns of the $(N-1)\times r$ matrix $\mat{W}$, also orthonormal, give the time coefficients of the directions.
The rank $r$ is the number of independent directions that the fit retains; those discarded are the directions along which the data scarcely vary, which are dominated by noise.

Second, the Moore--Penrose pseudo-inverse of the truncated matrix, the second expression of \cref{eq:svd}, solves \cref{eq:dmd} for $\mat{A}$ in the least-squares sense: $\mat{A} = \mat{V}_{2:N}\mat{V}_{1:N-1}^{+}$ maps the first data matrix as closely as possible onto the second, and of all the matrices that do so it is the one of smallest norm.

Third, the action of $\mat{A}$ is computed only on the $r$ directions of $\mat{U}$.
A snapshot in their span is $\snap = \mat{U}a$, with its $r$ coordinates in the vector $a$, and the fitted matrix advances these coordinates by the $r\times r$ matrix
\begin{equation}
  \widetilde{\mat{S}} = \mat{U}^{\herm}\mat{A}\,\mat{U} = \mat{U}^{\herm}\,\mat{V}_{2:N}\,\mat{W}\gmat{\Sigma}^{-1} ,
  \label{eq:Stilde}
\end{equation}
whose second form, obtained with $\mat{U}^{\herm}\mat{U} = \mat{I}$, contains only the data.

Fourth, the eigenvalues and eigenvectors of this small matrix,
\begin{equation}
  \widetilde{\mat{S}}\,z_l = \lambda_l\,z_l , \qquad l = 1,\dots,r ,
  \label{eq:dmd-eig}
\end{equation}
are the DMD eigenvalues $\lambda_l$ and the coordinates of the DMD modes $\mat{U}z_l$.

Why the eigenvalues of $\widetilde{\mat{S}}$, computed from the data alone, are eigenvalues of the unknown $\mat{A}$ is seen most easily when the snapshots come from a linear system with constant coefficients in which $r$ modes are excited.
Every snapshot is then a combination of the $r$ mode shapes $w_l$, $\snap_k = \sum_l a_l\lambda_l^{\,k}w_l$, so that, once there are at least $r$ snapshots, the columns of $\mat{V}_{1:N-1}$, and those of $\mat{U}$, span the same $r$-dimensional subspace as the mode shapes.
Because $\mat{A}w_l = \lambda_l w_l$, the matrix $\mat{A}$ maps that subspace into itself, so that $\mat{A}\mat{U} = \mat{U}\widetilde{\mat{S}}$ exactly, and if $\widetilde{\mat{S}}z = \lambda z$, then $\mat{A}(\mat{U}z) = \mat{U}\widetilde{\mat{S}}z = \lambda\,(\mat{U}z)$: the $r$ eigenvalues of $\widetilde{\mat{S}}$ are the $r$ eigenvalues of $\mat{A}$ that the record excites, and the vectors $\mat{U}z$ are their modes.
An eigenvalue of $\mat{A}$ whose mode is absent from the record leaves no trace in the data; DMD returns the part of the spectrum that the data contain.
The same holds when the snapshot is a linear measurement that determines the excited state \cite{Quinlan-2026-FD}, and with measured data, which contain noise and dynamics that no matrix of rank $r$ describes, the relations hold approximately.
The eigenvalue $\lambda_l$ is the factor by which mode $l$ is multiplied from one column to the next: its modulus gives the growth ($\abs{\lambda_l} > 1$) or decay ($\abs{\lambda_l}<1$) of the mode per step, and its argument the advance of its phase, so that its frequency is $\arg\lambda_l/(2\pi\dt)$.

\subsubsection{Systems with periodic coefficients}
\label{sec:floquet}

The matrix $\mat{A}$ exists only if the coefficients of the governing equations do not change with time.
The equations for small perturbations about a base flow have coefficients set by that base flow, and if the base flow oscillates periodically, so do the coefficients.
Consider therefore a linear system whose coefficients repeat with the period $T$,
\begin{equation}
  \dot{x} = \mat{L}(t)\,x \quad\text{with}\quad \mat{L}(t+T) = \mat{L}(t) ,
  \label{eq:ltp}
\end{equation}
where the state $x$ collects the perturbations of all the quantities of the flow and $\mat{L}(t)$ is the operator linearized about the periodic base state.
The pendulum on the oscillating pivot, \cref{eq:mathieu}, has this form, with $x = (\theta,\dot\theta)$ and $T = 2\pi/\omega_\mathrm{f}$.

The solution is still linear in the initial state, $x(t) = \gmat{\Phi}(t,t_0)\,x(t_0)$, where the state-transition matrix $\gmat{\Phi}(t,t_0)$ is the solution of $\partial_t\gmat{\Phi} = \mat{L}(t)\gmat{\Phi}$ with $\gmat{\Phi}(t_0,t_0) = \mat{I}$.
For constant coefficients it is the matrix exponential $\ee^{\mat{L}(t-t_0)}$, which depends only on the elapsed time; for periodic coefficients it depends on both times and is in general not the exponential of any simple matrix.
It has two properties that are used below.
Advancing from $t_0$ to $t_1$ and then from $t_1$ to $t$ is the same as advancing from $t_0$ to $t$, $\gmat{\Phi}(t,t_1)\gmat{\Phi}(t_1,t_0) = \gmat{\Phi}(t,t_0)$; and because the coefficients repeat, shifting both times by one period changes nothing, $\gmat{\Phi}(t+T,t_0+T) = \gmat{\Phi}(t,t_0)$.

By the second property the map over one whole period is the same for every period.
That map is the monodromy matrix,
\begin{equation}
  \mat{M} = \gmat{\Phi}(T,0) \quad\text{so that}\quad x(kT) = \mat{M}^{k}\,x(0) , \quad k = 0,1,2,\dots
  \label{eq:monodromy-ct}
\end{equation}
Viewed once per period, as under a stroboscope or on a Poincar\'e section, a system with periodic coefficients is a system with constant coefficients: one constant matrix advances the state from each period to the next.
The eigenvalues $\mu_l$ of $\mat{M}$ are the Floquet multipliers \cite{Floquet-1883,Hale-1980}.
A perturbation along the eigenvector $w_l$ of $\mat{M}$ is multiplied by $\mu_l$ in every period, so that it grows if $\abs{\mu_l} > 1$ and decays if $\abs{\mu_l} < 1$, and the system is unstable if any multiplier lies outside the unit circle.
Stability is decided by the multipliers and not by the eigenvalues of $\mat{L}(t)$ at any instant: for the pendulum with $\kappa > \abs{\epsilon}$ these are purely imaginary at every instant, which taken at face value would mean neutral oscillation, and yet within the tongues the pendulum is unstable.
Writing a multiplier as $\mu_l = \ee^{(\sigma_l+\ii\omega_l)T}$ defines a growth rate $\sigma_l = \ln\abs{\mu_l}/T$ and an angular frequency $\omega_l$ of the mode.
The growth rate is unique, but the frequency is determined only up to a multiple of $2\pi/T$, because a view once per period cannot count the whole cycles completed between two views.

Between the stroboscopic instants the solution has a simple structure.
Floquet's theorem \cite{Floquet-1883,Hale-1980} states that the state-transition matrix factors as $\gmat{\Phi}(t,0) = \mat{P}(t)\,\ee^{\mat{B}t}$, with $\mat{P}(t+T) = \mat{P}(t)$, $\mat{P}(0) = \mat{I}$, and $\mat{B}$ a constant matrix, so that $\mat{M} = \ee^{\mat{B}T}$.
Consequently the solution that starts along $w_l$ is
\begin{equation}
  x_l(t) = \gmat{\Phi}(t,0)\,w_l = \ee^{(\sigma_l+\ii\omega_l)t}\,\varphi_l(t) \quad\text{with}\quad \varphi_l(t+T) = \varphi_l(t) ,
  \label{eq:floquet-mode}
\end{equation}
a Floquet mode: a shape $\varphi_l(t)$ that changes periodically while the mode grows or decays at the single rate $\sigma_l$.
With constant coefficients the corresponding solution is $\ee^{(\sigma_l+\ii\omega_l)t}w_l$, whose shape is fixed.
Expanding the periodic shape in a Fourier series, $\varphi_l(t) = \sum_j \hat\varphi_{l,j}\,\ee^{2\pi\ii jt/T}$, shows that a Floquet mode contains the frequencies $\omega_l/2\pi + j/T$ for every harmonic $j$ present in it: in a spectrum it is not one line but a carrier with sidebands spaced by $1/T$, all of which grow or decay together.
Choosing another of the possible frequencies, $\omega_l + 2\pi n/T$, only renumbers the harmonics and leaves the solution unchanged: which line is called the carrier is a convention, fixed here by the dynamic pressure spectrum.
A fit with one constant matrix per sample has difficulty with such a system: DMD must represent one Floquet mode by several eigenvalues, one for each sideband, and when the sidebands are weak or buried in noise it may not represent the mode at all.

The data are recorded at the interval $\dt$, and the analysis period is taken to be a whole number $p$ of samples, $T = p\,\dt$.
With $x_k = x(k\dt)$ the samples obey
\begin{equation}
  x_{k+1} = \gmat{\Phi}_k\,x_k \quad\text{with}\quad \gmat{\Phi}_k = \gmat{\Phi}\bigl((k+1)\dt,\,k\dt\bigr) , \qquad \gmat{\Phi}_{k+p} = \gmat{\Phi}_k ,
  \label{eq:sampled}
\end{equation}
so that the matrix that advances the state by one sample changes from sample to sample and repeats after $p$ samples.
No constant matrix relates consecutive samples, but samples one period apart are related by the product of the $p$ matrices of one period,
\begin{equation}
  x_{k+p} = \mat{M}_c\,x_k \quad\text{with}\quad \mat{M}_c = \gmat{\Phi}_{c+p-1}\cdots \gmat{\Phi}_{c+1}\gmat{\Phi}_c ,
  \label{eq:monodromy}
\end{equation}
for every sample $k$ at phase $c$ of the period, that is, for $k = c,\,c+p,\,c+2p,\dots$, with $c = 1,\dots,p$.
The matrix $\mat{M}_c$ is the monodromy matrix with the period taken to begin at phase $c$.
The matrices of different phases are related by $\mat{M}_c = \gmat{\Phi}(c\dt,0)\,\mat{M}\,\gmat{\Phi}(c\dt,0)^{-1}$ and therefore have the same eigenvalues: the multipliers do not depend on the phase at which the period is taken to begin.
Their eigenvectors do: the eigenvector of $\mat{M}_c$ that belongs to $\mu_l$ is $\gmat{\Phi}(c\dt,0)\,w_l = \ee^{(\sigma_l+\ii\omega_l)c\dt}\varphi_l(c\dt)$, the Floquet mode at that phase.

The snapshot is not the state.
For small perturbations the measured quantities depend linearly on the state,
\begin{equation}
  \snap_k = \mat{C}_k\,x_k ,
  \label{eq:measurement}
\end{equation}
where the measurement matrix $\mat{C}_k$ says how each measured quantity depends on it.
A planar image of a three-dimensional flow and the dynamic pressure at three points see the state only in part, and $\mat{C}_k$ is never known.
The measurement may itself be periodic, $\mat{C}_{k+p} = \mat{C}_k$, as it is when a structure rotates through a fixed plane of measurement and presents a different section of itself at each phase.
A system with constant coefficients seen through a periodic measurement produces data of the same form as a periodic system \cite{Quinlan-2026-FD}, and what follows applies to both.
The methods use neither $\mat{C}_k$ nor $\gmat{\Phi}_k$; \cref{eq:sampled,eq:measurement} only state what the data are assumed to be.

\subsubsection{Floquet decomposition}

Schmid observed, in the paper that introduced DMD, that snapshots sampled at the period of an oscillatory flow would identify the Floquet matrix \cite[\S 2.6]{Schmid-2010}.
FD and EFD develop that observation into methods, and the companion paper gives their analysis \cite{Quinlan-2026-FD}.
FD is DMD applied to the record subsampled once per period, starting at phase $c$.
If $K$ is the number of whole periods in the record, its data matrix is
\begin{equation}
  \mat{V}^{(c)}_{1:K} = \begin{bmatrix} \snap_c & \snap_{c+p} & \snap_{c+2p} & \cdots & \snap_{c+(K-1)p} \end{bmatrix} ,
  \label{eq:fd}
\end{equation}
and $\mat{V}^{(c)}_{1:K-1}$ and $\mat{V}^{(c)}_{2:K}$, its first and last $K-1$ columns, take the places of $\mat{V}_{1:N-1}$ and $\mat{V}_{2:N}$ in \cref{eq:dmd,eq:svd,eq:Stilde,eq:dmd-eig}.
By \cref{eq:monodromy,eq:measurement} the columns are $\snap_{c+(k-1)p} = \mat{C}_c\,\mat{M}_c^{\,k-1}x_c$, successive iterates of the monodromy matrix seen through the measurement at phase $c$.
If the snapshot were the whole state, they would satisfy $\mat{V}^{(c)}_{2:K} = \mat{M}_c\mat{V}^{(c)}_{1:K-1}$, which is \cref{eq:dmd} with the monodromy matrix in the place of $\mat{A}$, and the four steps of DMD would return the excited multipliers as the eigenvalues $\muh_l$ of $\widetilde{\mat{S}}$ and the eigenvectors of $\mat{M}_c$ as modes.
The same is true when the measurement at one phase determines the excited state: the snapshot then determines the state, the state determines the snapshot one period later, and a constant matrix relates the snapshots one period apart, with the excited multipliers as its eigenvalues and the measured Floquet modes $\mat{C}_c\varphi_l(c\dt)$ as its modes \cite{Quinlan-2026-FD}.
The record is taken at the rate that the hardware allows, and the analysis period is chosen afterward.

The condition on the measurement can fail, and for the present data it does.
One displacement signal of an oscillator cannot determine both its displacement and its velocity, and a planar image of a three-dimensional flow and the dynamic pressure at three points do not determine the state either, so that FD returns eigenvalues that need not be multipliers.
For measurements that depend nonlinearly on the state, the matrix that DMD fits approximates the Koopman operator of the map over one period \cite{Rowley-2009,Mezic-2016}, under the same requirement.
FD also discards all but one of the $p$ snapshots of each period.

\subsubsection{Ensemble Floquet decomposition}

EFD removes both limitations by stacking the $p$ snapshots of each period into one tall column and applying DMD to columns one period apart.
The stacked snapshot of period $k$ and the stacked data matrix are
\begin{equation}
  \tilde\snap_k = \begin{bmatrix} \snap_{(k-1)p+1} \\ \snap_{(k-1)p+2} \\ \vdots \\ \snap_{kp} \end{bmatrix},
  \qquad
  \widetilde{\mat{V}}_{1:K} = \begin{bmatrix} \tilde\snap_1 & \tilde\snap_2 & \cdots & \tilde\snap_K \end{bmatrix}
  = \begin{bmatrix} \mat{V}^{(1)}_{1:K} \\ \mat{V}^{(2)}_{1:K} \\ \vdots \\ \mat{V}^{(p)}_{1:K} \end{bmatrix} ,
  \label{eq:efd}
\end{equation}
so that block row $c$ of the stacked matrix is the FD matrix of phase $c$ (\cref{fig:matrices}).
By \cref{eq:sampled,eq:measurement} the stacked snapshot is a linear image of the state at the start of its period,
\begin{equation}
  \tilde\snap_k = \mat{O}\,x_{(k-1)p+1} \quad\text{with}\quad
  \mat{O} = \begin{bmatrix} \mat{C}_1 \\ \mat{C}_2\gmat{\Phi}_1 \\ \mat{C}_3\gmat{\Phi}_2\gmat{\Phi}_1 \\ \vdots \\ \mat{C}_p\gmat{\Phi}_{p-1}\cdots\gmat{\Phi}_1 \end{bmatrix}
  \quad\text{so that}\quad
  \tilde\snap_{k+1} = \mat{O}\,\mat{M}_1\,x_{(k-1)p+1} .
  \label{eq:efd-obs}
\end{equation}
Block $c$ of the matrix $\mat{O}$ carries the state from the first phase of the period to phase $c$ and measures it there.
It is the observability matrix of the system over one period, and it is the same for every period because the dynamics and the measurement repeat.
The stacked snapshots are therefore successive iterates of the monodromy matrix $\mat{M}_1$ seen through the constant matrix $\mat{O}$, the situation of FD with $\mat{O}$ in the place of $\mat{C}_c$.
The argument given there applies unchanged, with a different requirement: EFD returns the excited multipliers exactly when $\mat{O}$ determines the excited state, that is, when the measurements made during one period determine the state at its start \cite{Quinlan-2026-FD}.
The four steps of DMD, applied to $\widetilde{\mat{V}}_{1:K-1}$ and $\widetilde{\mat{V}}_{2:K}$, give the EFD: the eigenvalues $\muh_l$ of $\widetilde{\mat{S}}$ estimate the multipliers, and each mode $\tilde\psi_l = \mat{U}z_l$ is a stack of $p$ blocks, $\tilde\psi_l^{[1]},\dots,\tilde\psi_l^{[p]}$, one for each phase of the period.
The mode is $\mat{O}$ times the eigenvector of $\mat{M}_1$, so that by \cref{eq:floquet-mode} its block $c$ is, up to a common factor,
\begin{equation}
  \tilde\psi_l^{[c]} = \mat{C}_c\,\gmat{\Phi}(c\dt,0)\,w_l = \ee^{(\sigma_l+\ii\omega_l)c\dt}\;\mat{C}_c\,\varphi_l(c\dt) ,
  \label{eq:efd-block}
\end{equation}
the measured Floquet mode at phase $c$, multiplied by its growth and its advance in phase over the fraction $c/p$ of the period.
Every multiplier appears in all $p$ FD problems but only once in the result, because all the blocks of a stacked snapshot are determined by the state at the start of its period.
The ensemble in the name is the set of the $p$ period-sampled sequences contained in one record, not an ensemble of experiments; estimating them together gives every phase the same multipliers and uses every snapshot.

Stacking thus changes what the measurement must provide.
FD requires the measurement at one instant to determine the state; EFD requires only that the measurements made during one period determine it.
The second is a much weaker demand: what cannot be seen at one phase is carried into view at another, as the velocity of an oscillator is determined by its displacement over an interval, and a structure that rotates through a fixed plane shows one section of itself in each image and all of them over one revolution.
For a linear periodic system the companion paper proves that FD is exact under the first condition and EFD under the second \cite{Quinlan-2026-FD}, and that difference decides the outcome for the present data (\cref{sec:res-efd}).

\subsubsection{Computation}

The computation, given as an algorithm in the companion paper \cite{Quinlan-2026-FD}, takes five steps.
\begin{enumerate}
\item Form the composite snapshots and scale each kind of data (\cref{sec:snapshots}); choose the number $p$ of samples in the analysis period and the rank $r$.
\item Arrange the record into the stacked matrix $\widetilde{\mat{V}}_{1:K}$ of \cref{eq:efd}, discarding the samples of an incomplete last period.
\item Compute the truncated SVD of $\widetilde{\mat{V}}_{1:K-1}$, \cref{eq:svd}. When the stacked snapshots have many more entries than there are columns, as here (about $1.7\times10^{5}$ entries and $850$ columns), the SVD is obtained from the eigenvalues and eigenvectors of the $(K-1)\times(K-1)$ matrix of the inner products of the columns, $\widetilde{\mat{V}}_{1:K-1}^{\herm}\widetilde{\mat{V}}_{1:K-1} = \mat{W}\gmat{\Sigma}^2\mat{W}^{\herm}$, whose entries are sums over the $p$ phases of the inner products of single snapshots, so that the stacked matrix need not be formed.
\item Form $\widetilde{\mat{S}}$ by \cref{eq:Stilde}, with $\widetilde{\mat{V}}_{2:K}$ in the place of $\mat{V}_{2:N}$, and compute its eigenvalues $\muh_l$ and eigenvectors $z_l$, \cref{eq:dmd-eig}.
\item Compute the modes, $\tilde\psi_l = \mat{U}z_l = \widetilde{\mat{V}}_{1:K-1}\mat{W}\gmat{\Sigma}^{-1}z_l$, split each into its $p$ blocks, and obtain the frequencies and harmonics as described next.
\end{enumerate}
FD is the same computation with a single block row, and DMD is FD with $p = 1$.

\subsubsection{Frequencies, modes, and harmonics}

The frequency of mode $l$ is $\arg(\muh_l)/(2\pi T)$ up to a multiple of $1/T$, which the dynamic pressure spectrum fixes, since a view once per period cannot count the whole cycles completed in between.
The analysis period is that of the periodicity presumed in the dynamics, not that of a mode, and a system whose coefficients do not change with time is periodic with any period: EFD with any $T$ returns each of its modes, of growth rate $\sigma$ and angular frequency $\omega$, as $\mu = \ee^{(\sigma+\ii\omega)T}$, so the identification of a mode is not by itself evidence that the dynamics are periodic.

By \cref{eq:efd-block}, block $c$ of an EFD mode carries the fraction $c/p$ of the change $\muh_l$ over the period; removing it leaves the measured periodic shape of the mode at phase $c$, and a discrete Fourier transform across the aligned blocks gives the harmonics of that shape over the period,
\begin{equation}
  \hat\psi_{l,j} = \frac{1}{p}\sum_{c=1}^{p} \ee^{-2\pi\ii\,jc/p}\;\muh_l^{-c/p}\;\tilde\psi_l^{[c]}, \qquad j = 0,\dots,p-1 ,
  \label{eq:harmonics}
\end{equation}
where the principal value of the root is taken and harmonics $j$ and $j-p$ are the same.
Harmonic zero, the average of the aligned blocks, is the mode shown in this paper, and the averaging suppresses what is not repeated from phase to phase.
Harmonic $j$ holds the content of the mode at its frequency plus $j/T$, the sidebands of a Floquet mode (\cref{sec:floquet}); a mode whose largest harmonic is not harmonic zero is the alias of content at that higher frequency, and \cref{sec:res-param} uses the harmonics to test for a coupling between the instability and the faster motion that sets the analysis period.

\subsubsection{Relation to DMD and Hankel-DMD}

\begin{figure}[tb]
\centering
\newcommand{\cellborders}[3]{\foreach \i in {1,...,#2}{\foreach \j in {1,...,#3}{\draw[line width=0.3pt] (#1-\i-\j.north west) rectangle (#1-\i-\j.south east);}}}
\begin{tikzpicture}[font=\scriptsize,
  cell/.style={minimum width=6.3mm, minimum height=4.4mm, inner sep=0pt, outer sep=0pt, anchor=center, text height=1.2ex, text depth=0.45ex},
  ph0/.style={fill=white}, ph1/.style={fill=black!14}, ph2/.style={fill=black!36},
  grid/.style={matrix of math nodes, nodes={cell}, column sep=0pt, row sep=0pt, inner sep=0pt},
  tstep/.style={-{Stealth[length=3.5pt]}, line width=0.4pt}]
  \matrix (rec) [grid] at (0,0) {
    |[ph0]| \snap_1 & |[ph1]| \snap_2 & |[ph2]| \snap_3 & |[ph0]| \snap_4 & |[ph1]| \snap_5 & |[ph2]| \snap_6 &
    |[ph0]| \snap_7 & |[ph1]| \snap_8 & |[ph2]| \snap_9 & |[ph0]| \snap_{10} & |[ph1]| \snap_{11} & |[ph2]| \snap_{12} \\};
  \cellborders{rec}{1}{12}
  \node[right=0.5mm of rec] {$\cdots$};
  \node[left=1.5mm of rec, font=\small] {a)};
  \draw[decorate, decoration={brace, amplitude=3pt}] ([yshift=2pt]rec-1-1.north west) -- node[above=3pt] {one period $T$, $p=3$} ([yshift=2pt]rec-1-3.north east);
  \draw[decorate, decoration={brace, amplitude=3pt, mirror}] ([yshift=-2pt]rec-1-1.south west) -- node[below=3pt] {$\dt$} ([yshift=-2pt]rec-1-1.south east);
  \node[font=\small] at (-3.35,-1.45) {shift $s=1$};
  \node[font=\small] at (3.35,-1.45) {shift $s=p$};
  \matrix (dmd) [grid] at (-3.35,-2.45) {
    |[ph0]| \snap_1 & |[ph1]| \snap_2 & |[ph2]| \snap_3 & |[ph0]| \snap_4 & |[ph1]| \snap_5 \\};
  \cellborders{dmd}{1}{5}
  \node[right=0.5mm of dmd] {$\cdots$};
  \node[above=0.6mm of dmd, font=\small] {b) DMD};
  \draw[tstep] ([yshift=-1.5pt]dmd-1-1.south) to[bend right=50] node[below=0.5pt] {$\dt$} ([yshift=-1.5pt]dmd-1-2.south);
  \matrix (fd) [grid] at (3.35,-2.45) {
    |[ph0]| \snap_1 & |[ph0]| \snap_4 & |[ph0]| \snap_7 & |[ph0]| \snap_{10} \\};
  \cellborders{fd}{1}{4}
  \node[right=0.5mm of fd] {$\cdots$};
  \node[above=0.6mm of fd, font=\small] {c) FD, phase $c=1$};
  \draw[tstep] ([yshift=-1.5pt]fd-1-1.south) to[bend right=50] node[below=0.5pt] {$T$} ([yshift=-1.5pt]fd-1-2.south);
  \matrix (hk) [grid] at (-3.35,-4.75) {
    |[ph0]| \snap_1 & |[ph1]| \snap_2 & |[ph2]| \snap_3 & |[ph0]| \snap_4 & |[ph1]| \snap_5 \\
    |[ph1]| \snap_2 & |[ph2]| \snap_3 & |[ph0]| \snap_4 & |[ph1]| \snap_5 & |[ph2]| \snap_6 \\
    |[ph2]| \snap_3 & |[ph0]| \snap_4 & |[ph1]| \snap_5 & |[ph2]| \snap_6 & |[ph0]| \snap_7 \\};
  \cellborders{hk}{3}{5}
  \node[right=0.5mm of hk] {$\cdots$};
  \node[above=0.6mm of hk, font=\small] {d) Hankel-DMD};
  \draw[tstep] ([yshift=-1.5pt]hk-3-1.south) to[bend right=50] node[below=0.5pt] {$\dt$} ([yshift=-1.5pt]hk-3-2.south);
  \matrix (efd) [grid] at (3.35,-4.75) {
    |[ph0]| \snap_1 & |[ph0]| \snap_4 & |[ph0]| \snap_7 & |[ph0]| \snap_{10} \\
    |[ph1]| \snap_2 & |[ph1]| \snap_5 & |[ph1]| \snap_8 & |[ph1]| \snap_{11} \\
    |[ph2]| \snap_3 & |[ph2]| \snap_6 & |[ph2]| \snap_9 & |[ph2]| \snap_{12} \\};
  \cellborders{efd}{3}{4}
  \node[right=0.5mm of efd] {$\cdots$};
  \node[above=0.6mm of efd, font=\small] {e) EFD};
  \draw[tstep] ([yshift=-1.5pt]efd-3-1.south) to[bend right=50] node[below=0.5pt] {$T$} ([yshift=-1.5pt]efd-3-2.south);
  \node[rotate=90, font=\small] at (-6.3,-2.45) {$d=1$};
  \node[rotate=90, font=\small] at (-6.3,-4.75) {$d=p$};
\end{tikzpicture}
\caption{Data matrices of the four decompositions, built from one record and drawn for $p=3$ samples per period (after \cite{Quinlan-2026-FD}); the shading marks the phase within the period. a) The record of snapshots. b) Standard DMD, $\mat{V}_{1:N}$. c) FD, $\mat{V}^{(1)}_{1:K}$: one snapshot per period. d) Hankel-DMD: $p$ consecutive snapshots stacked, shifted by one sample. e) EFD, $\widetilde{\mat{V}}_{1:K}$: the same stack, shifted by one period, so that each block row is the FD matrix of one phase. DMD fits the matrix that maps each column to the next; the arrows give the time between columns.}
\label{fig:matrices}
\end{figure}
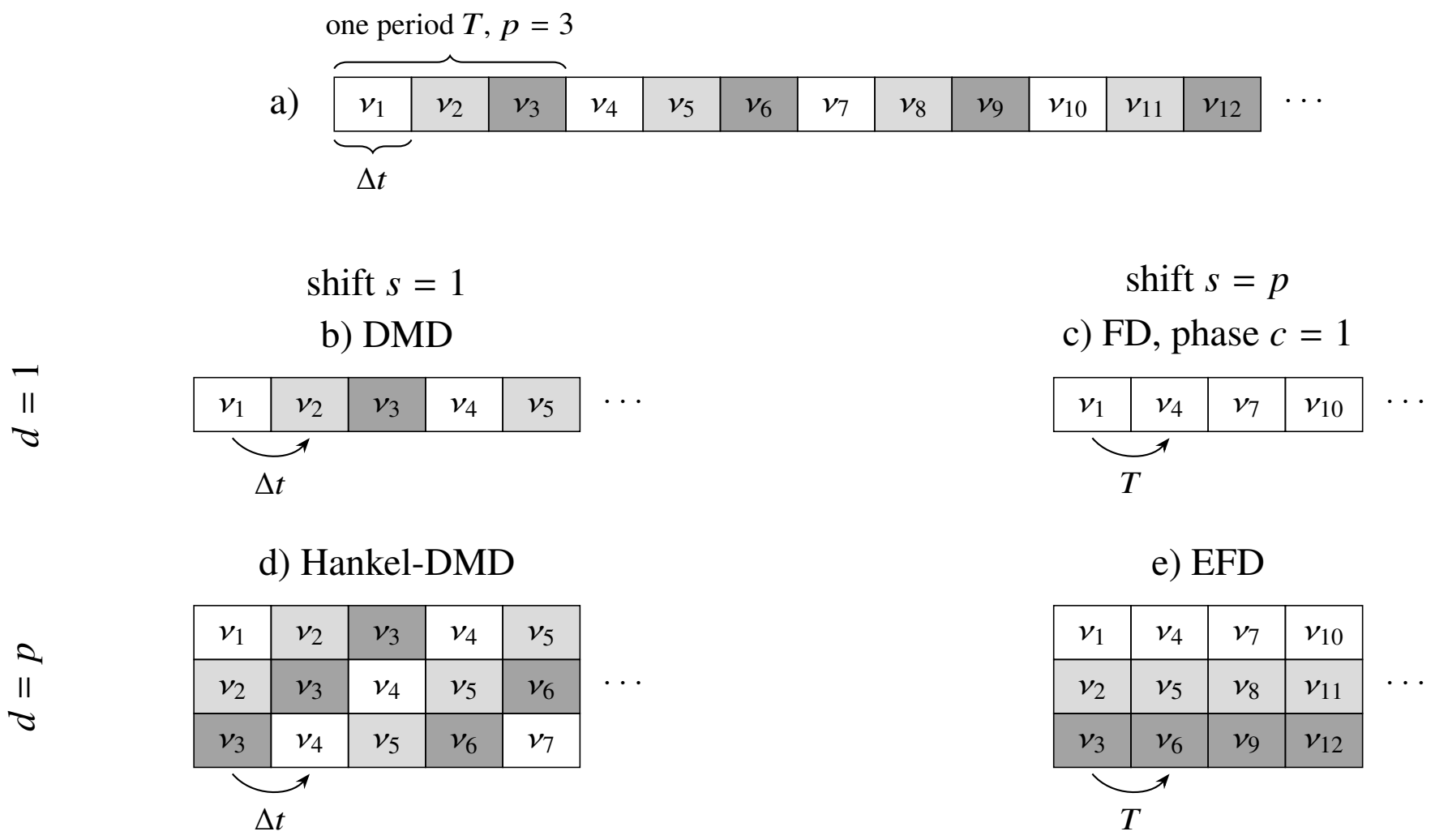

The four decompositions compared in this paper are one computation, \cref{eq:svd,eq:Stilde,eq:dmd-eig}, applied to data matrices that differ in the number $d$ of consecutive snapshots stacked in each column and in the shift $s$ between columns (\cref{fig:matrices}): standard DMD has $(d,s) = (1,1)$, FD $(1,p)$, and EFD $(p,p)$, and Hankel-DMD \cite{Arbabi-2017,LeClainche-2017} stacks the same $p$ snapshots as EFD but shifts by one sample, $(d,s) = (p,1)$.
The shift is the time step of the matrix that is fitted, and the two methods with a window come from different views of the data.
Hankel-DMD keeps the view of standard DMD, a system with constant coefficients advanced by one sample, and stacks consecutive snapshots to supply the parts of the state that a single snapshot does not measure, so that the matrix it fits approximates the Koopman operator over one sample \cite{Arbabi-2017}.
EFD takes the view of Floquet theory, a system with periodic coefficients or a periodic measurement advanced by one period, and stacks the snapshots of one period because together they determine the state at its start (\cref{eq:efd-obs}).
For a linear periodic system the eigenvalues $\beta$ of Hankel-DMD are $p$th roots of the Floquet multipliers, $\beta^{\,p} = \mu$, one for each harmonic present in the mode, and the first block of the mode returned with each is the corresponding harmonic of \cref{eq:harmonics} \cite{Quinlan-2026-FD}; below, the estimates $\betah^{\,p}$ are compared with $\muh$.
With a window of one period the two methods hold the same information, but EFD returns the multipliers directly, needs one eigenvalue for each mode instead of one for each harmonic, and resolves each mode in the phase of the period.
Hankel-DMD is used here with that window, $d = p$; on the present data shorter windows return a mixture of the two modes, because the window must span about one cycle of the hydrodynamic mode near $1$\,kHz to determine enough of the state \cite{Quinlan-2026-FD}.
Knowing the period therefore gives the user of Hankel-DMD its shortest window, $d \ge p$; for EFD it is the shift itself.

\subsection{Spectral estimates}
\label{sec:spectral}

Spectra, cross-spectra, and coherences are Welch estimates with Hann windows and $75\%$ overlap \cite{Bendat-2010}, with segments of $1024$ samples for the $10$\,kHz records and longer ones for the $4$\,s dynamic pressure records; the cross-spectrum $S_{ab} = \langle \hat a\,\hat b^* \rangle$ has the phase by which $a$ leads $b$.
The dynamic pressure spectra use the records as acquired.
Where the dynamic pressure is compared with the optical records, it is taken at the instants of the frames (\cref{sec:diagnostics}), band-passed between $80$ and $500$\,Hz at the frequencies of the duct modes below $500$\,Hz, and as its time derivative, which has the same coherence, wherever higher frequencies are involved.

The heat-release fluctuation $q'$ is taken proportional to the \chstar\ signal, an uncalibrated approximation to the heat release \cite{Lauer-2010}, so only its coherence, its phase, and comparisons within one record are used, through three quantities.
The normalized Rayleigh index at frequency $f$,
\begin{equation}
  \mathcal{R}(f) = \frac{\operatorname{Re} S_{qp}}{\sqrt{S_{qq}S_{pp}}} = \gamma_{qp}\cos\varphi_{qp} ,
  \label{eq:rayleigh}
\end{equation}
is the product of the coherence and the cosine of the phase between heat release and acoustic pressure, positive when the heat release adds energy to the acoustic field \cite{Rayleigh-1878}; the local Rayleigh density is its numerator for each pixel of the \chstar\ image, summed over the band of the mode; and the response per unit acoustic pressure is $H_{qp} = S_{qp}/S_{pp}$, whose real part is the heat release in phase with the acoustic pressure.

Two tests look for a coupling between a slow oscillation at frequency $f_1$ and faster activity.
The bicoherence \cite{Kim-1979},
\begin{equation}
  b^2(f_1,f_2) = \frac{\abs{\langle \hat a(f_1)\,\hat b(f_2)\,\hat c^*(f_1+f_2) \rangle}^2}{\langle \abs{\hat a(f_1)\,\hat b(f_2)}^2 \rangle \, \langle \abs{\hat c(f_1+f_2)}^2 \rangle} ,
  \label{eq:bicoherence}
\end{equation}
measures whether the phases of three components keep a fixed sum from one segment of the record to the next, as they do when the components are coupled quadratically or when the oscillation at $f_1$ modulates the one at $f_2$.
It lies between $0$ and $1$, and a linear filter applied to a signal does not change it, so the standoff tubes do not affect it.
It is computed from the $4$\,s dynamic pressure records in segments of $0.1$\,s, and its $95\%$ level is that of its values when the factor at $f_1$ is taken from another segment, which breaks any relation between phases at the same instant.
The second test is the coherence with the dynamic pressure at $f_1$ of the envelope of the signal band-passed between $800$ and $1200$\,Hz, which detects a modulation of activity whose frequency wanders but is diluted by the pairs of frequencies that are not coupled; its $95\%$ level is its value at neighboring frequencies.

\subsection{Acoustic loss through the exhaust}
\label{sec:loss-model}

The eight exhaust holes are the only outlets of the duct, and the mean flow through them makes them absorb sound \cite{Bechert-1980,Howe-1979}.
At the frequencies of interest the flow through them is quasi-steady, with a Strouhal number $\omega d_h/U_j$ near $0.05$ and a Mach number near $0.3$, so that, with $\widebar{P} - P_a = \rho U_j^2/2$, an acoustic pressure $p'_e$ at the holes changes the jet velocity by $u'_j = p'_e/(\rho U_j)$, and the holes remove the period-averaged power
\begin{equation}
  W = \frac{A_j^2}{2\,\dot m}\,\abs{p_e}^2 \quad\text{with}\quad A_j = C_d A_h ,
  \label{eq:W}
\end{equation}
where $A_h$ is their total area, $\dot m = \rho U_j A_j$ the mass flow, and $\abs{p_e}$ the acoustic pressure amplitude; the holes act as a resistance that does not depend on the exhaust temperature.
With the period-averaged energy $E = A_d\int\abs{p}^2\,\mathrm{d}x/(2\gamma\widebar{P})$ of a one-dimensional mode of acoustic pressure amplitude $\abs{p(x)}$ in a duct of area $A_d$, the exhaust alone makes the amplitude decay at the rate
\begin{equation}
  \alpha_e = \frac{W}{2E} = \frac{\gamma\widebar{P} A_j^2}{2\,\dot m A_d}\;\frac{\langle\abs{p_e}^2\rangle}{\int\abs{p}^2\,\mathrm{d}x} ,
  \label{eq:alpha}
\end{equation}
where the average is over the positions of the holes.
The loss is proportional to the square of the acoustic pressure at the exhaust, so a mode with a pressure node there is not damped by it, and a stable mode driven by broadband noise has a half-power bandwidth $\alpha/\pi$, where $\alpha$ is its net decay rate, which is how \cref{eq:alpha} is compared with the measured spectra.
The evaluation uses the eight holes of $6.35$\,mm diameter in the $12.7$\,mm upper wall, at $x = \pm 12.7$ and $\pm 38.1$\,mm, the measured mass flow, a discharge coefficient between $0.6$ and $0.85$, and the mode shapes of a closed duct with either a uniform sound speed or the temperature profile of \cite{Quinlan-2015}.
The exhaust was drawn off above the holes by an ejector, so the pressure difference across them was not measured and the absolute loss rates are rough estimates; the argument of \cref{sec:res-why} uses only the ratio of the losses of different modes, which depends on their shapes.

%% file: sections/04_results.tex
\section{Results}
\label{sec:results}

\subsection{Decomposition of the two configurations}
\label{sec:res-efd}

The analysis period is $T = 1$\,ms, $p = 10$ samples, the period of the narrowband mode of the injector shear layer near $1$\,kHz \cite{Quinlan-2017} examined in \cref{sec:res-khz}; it is not a sharp period, and the results do not depend on it within $\pm10\%$.
The rank is $r = 25$.

\Cref{tab:methods} lists what the four decompositions return near the two lowest peaks of the dynamic pressure spectrum of the unstable configuration \cite{Quinlan-2026-FD}.
Standard DMD returns nothing near either frequency, in agreement with \cite{Quinlan-2015}, and neither does FD at any of the ten phases of the period.
EFD returns both, close to the unit circle, and Hankel-DMD returns the same two pairs, with modes that agree closely with the phase-averaged EFD modes.
What separates the two methods that find the instability from the two that do not is the stacking of consecutive snapshots (the lower row of \cref{fig:matrices}): the measurements at one instant do not determine the state of the flow, and those over one analysis period do so well enough.

\begin{table}[tb]
\caption{Eigenvalue of largest modulus within 8\,Hz of the two lowest peaks of the dynamic pressure, unstable configuration, baseline weights, $p=10$, $r=25$ \cite{Quinlan-2026-FD}. Moduli are per analysis period; a dash means that none exceeds 0.25.}
\label{tab:methods}
\centering
\begin{tabular}{lcccc}
\toprule
 & \multicolumn{2}{c}{First pair} & \multicolumn{2}{c}{Second pair} \\
\cmidrule(lr){2-3}\cmidrule(lr){4-5}
Method & $f$, Hz & Modulus & $f$, Hz & Modulus \\
\midrule
DMD & --- & --- & --- & --- \\
FD, each of ten phases & --- & --- & --- & --- \\
Hankel-DMD & $167.74$ & $0.9209$ & $263.70$ & $0.9726$ \\
EFD & $171.32$ & $0.9530$ & $262.48$ & $0.9524$ \\
\bottomrule
\end{tabular}
\end{table}

\begin{figure}[tb]
\centering
\includegraphics[width=\textwidth]{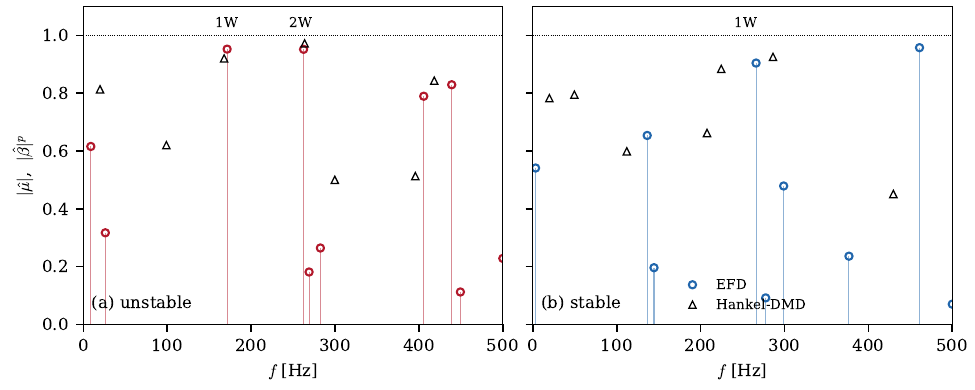}
\caption{Moduli $\abs{\muh}$ of the EFD multipliers and $\abs{\betah}^{\,p}$ of the Hankel-DMD eigenvalues raised to the power $p$, against frequency, for the baseline settings. a) Unstable configuration. b) Stable configuration. The labels mark the duct modes. }
\label{fig:efd}
\end{figure}

\Cref{fig:efd} compares the two configurations.
In the unstable one the two pairs have the largest moduli of all the oscillatory eigenvalues; the next, at $439$\,Hz, is the third duct mode, and a pair at $406$\,Hz is an alias of the fourth duct mode near $594$\,Hz, which the $1$\,ms period folds back (\cref{eq:harmonics}).
In the stable configuration EFD returns a pair at $266$\,Hz with the signature of the \oneW\ of the shorter duct, the injector signals equal and in antiphase with the duct signal, which Hankel-DMD does not return, and neither method has a pair near $170$\,Hz.
EFD also returns a weaker pair at $136$\,Hz, where the duct has no mode (\cref{sec:res-why}), and its strongest pair, at $461$\,Hz, is again an alias, of the second duct mode of the shorter duct.
The values differ slightly from those of \cite{Quinlan-2015,Quinlan-2022}, which were obtained with differently filtered dynamic pressure signals, before the dynamic pressure and optical records were aligned (\cref{sec:diagnostics}), and with the commercial velocity vectors.

The identification is robust within limits, which the companion paper tabulates \cite{Quinlan-2026-FD}.
The two pairs are found whenever the dynamic pressure is weighted at least three times as heavily as the scalar data and the rank is at least $15$, including in each half of the record analyzed separately; with a weaker weight on the dynamic pressure they are weakened or lost.
They are found also when the analysis period is changed, by one sample either way or, with the record resampled, to any period between $1.0$ and $1.3$\,ms, and the frequency of the first pair then changes by about $1\%$.
They depend on the band of the dynamic pressure signals: as its low-pass cutoff is raised, DMD and FD lose the $170$\,Hz mode above about $400$\,Hz and EFD above about $660$\,Hz \cite{Quinlan-2026-FD}, as the higher duct modes and the standoff resonance enter the dynamic pressure.
The modes are therefore identified through the band-limited dynamic pressure, and their velocity and scalar parts are the parts of those fields that move with the dynamic pressure.
These parts are small in the scaled snapshot, in which the dynamic pressure entries carry more than $99\%$ of the squared norm of each mode, but they are what \cref{sec:res-mode,sec:res-cl} analyze, and a Rayleigh analysis asks for precisely the part of the heat release that is coherent with the acoustic pressure.

The moduli are below one, yet the instability does not decay.
Projected on the composite snapshots, the coefficient of the $170$\,Hz mode keeps a steady mean amplitude and a nearly constant phase over the $110$ cycles after the first $0.2$\,s, whereas a modulus of $0.953$ per millisecond, taken literally, would make it decay by a factor of $\ee$ in $21$\,ms; the $262$\,Hz mode and the modes of the stable configuration come in bursts with wandering phases.
The oscillation at $170$\,Hz is a limit cycle, whose modulus is one, and its estimate falls below one because noise and the wandering of the phase reduce the correlation from one period to the next, a bias toward zero known for DMD eigenvalues \cite{Dawson-2016,Hemati-2017}.
We therefore read the moduli as measures of the coherence of a mode from one period to the next, which order the modes robustly, and not as decay rates; that the $170$\,Hz mode is unstable rests on the Rayleigh analysis of \cref{sec:res-cl}.

\subsection{The unstable mode is the first transverse mode}
\label{sec:res-mode}

The dynamic pressure parts of the modes identify them (\cref{tab:modes}).
In the $170$\,Hz mode the three dynamic pressure signals are nearly in phase, and the injector signals are a fifth to a quarter of the duct signal.
In a one-dimensional duct the acoustic pressure amplitude of the first mode grows monotonically from its single node toward each end, whatever the distributions of sound speed and area, so the node of the \oneW\ lies beyond the injector, on the side away from the sensing section, about $0.14$\,m from it for the temperature distribution of \cite{Quinlan-2015}, and the injector and the exhaust above it are near the pressure node and velocity antinode of the mode.
In the $262$\,Hz mode the injector signals are nearly equal, larger than the duct signal and in antiphase with it: this is the \twoW, with the injector near a pressure antinode, and the \oneW\ of the stable configuration, at $254$\,Hz, has the same signature.

\begin{figure}[tb]
\centering
\includegraphics[width=\textwidth]{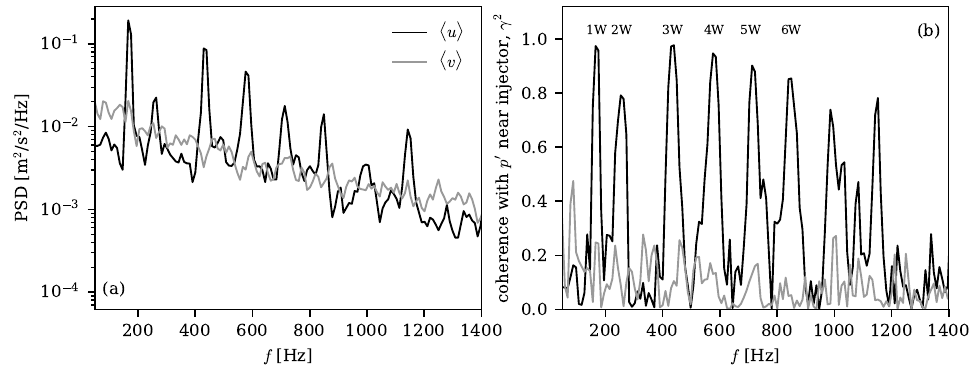}
\caption{Velocity averaged over the points of the image at which at least 80\% of the vectors are accepted, unstable configuration. a) Power spectral density of the transverse ($u$) and axial ($v$) components. b) Coherence with the combustor dynamic pressure near the injector; the labels mark the duct modes.}
\label{fig:velocity}
\end{figure}

The acoustic velocity at the injector is measured by the PIV.
The transverse velocity averaged over the image has a coherence of $0.98$ at the \oneW\ with the combustor dynamic pressure near the injector (\cref{fig:velocity}), with a phase nearly uniform over the field, as expected of an acoustic wave for which the field of view is compact, and its coherence shows the whole ladder of duct modes.
The amplitude of the acoustic velocity at the \oneW\ is $2.1$ to $2.6$\,m/s, depending on which vectors are averaged.
It leads the combustor dynamic pressure near the injector by $100^\circ$, and with the difference in acoustic pressure between the two injector sensors the momentum equation gives a gas density of $0.37$ to $0.50$\,kg/m$^3$, that of gas at $700$ to $1000$\,K, so the sensitivity of the dynamic pressure measurement (\cref{sec:diagnostics}) is consistent with the velocity measurement.

\begin{figure}[tb]
\centering
\includegraphics[width=\textwidth]{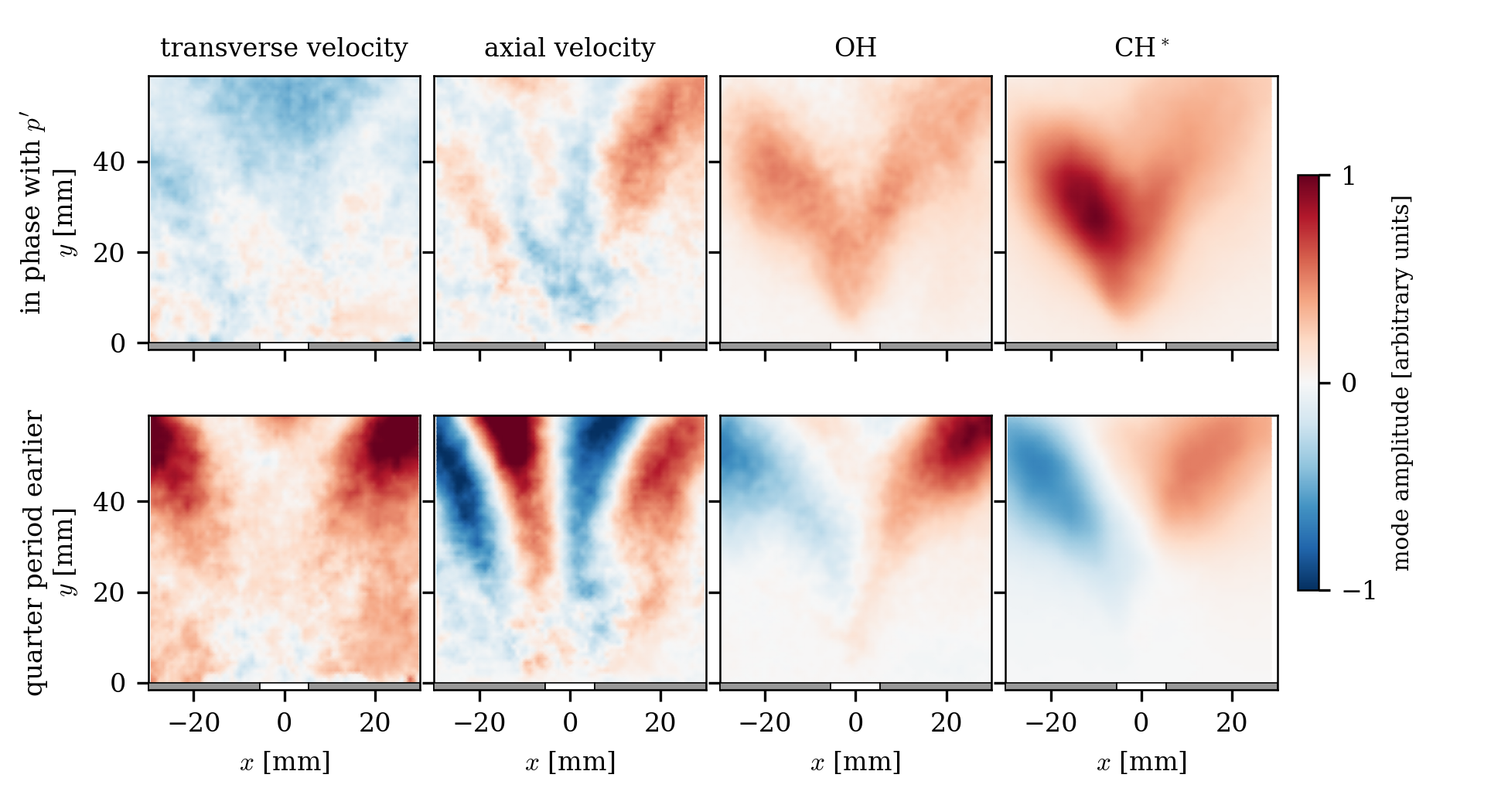}
\caption{The EFD mode at $170$\,Hz, unstable configuration: average over the ten phases of the analysis period. Top row: the part in phase with the combustor dynamic pressure near the injector. Bottom row: the part that leads it by a quarter period. Each field is scaled by its own maximum. The injector face and nozzle are drawn below each panel; positive $x$, toward the sensing section, is to the right.}
\label{fig:mode170}
\end{figure}

\Cref{fig:mode170} shows the optical parts of the $170$\,Hz mode, with phase referenced to the combustor dynamic pressure near the injector.
The transverse velocity is almost entirely in the lower row, a quarter period ahead of the acoustic pressure, and has one sign over the field.
This is the bulk transverse velocity in phase with the acoustic velocity that was noted in \cite{Quinlan-2022}; it is the acoustic velocity of the \oneW\ itself.
The axial velocity shows bands of alternating sign along the two branches of the swirling jet, also mainly a quarter period ahead of the acoustic pressure, which strengthen downstream toward the exhaust.
They are the signature of the jets and their shear layers being displaced sideways by the acoustic motion, and of the vortical structure that develops from that displacement downstream of the injector \cite{Quinlan-2022}.
Averaged over the field, the axial velocity is coherent with the acoustic pressure mainly on the branch toward the sensing section, where the coherence is $0.58$ against $0.04$ for the other half of the field.
The OH and \chstar\ parts have a real component in phase with the acoustic pressure that has one sign over the whole flame, and an imaginary component that changes sign across the axis; \cref{sec:res-cl} quantifies both.

\subsection{Why the first transverse mode of the long duct is the unstable one}
\label{sec:res-why}

The two configurations differ only in the length of the duct, and with it in the frequencies of the modes and in where the injector sits in each.
Three properties single out the \oneW\ of the long duct.

The first is the way the mode couples to the flame.
An acoustic mode can act on an injector flame in two ways.
Its acoustic pressure can modulate the flow of reactants through the injector, and its velocity can move the flow and the flame.
This injector was designed with a large pressure drop, more than $10$\,bar at these conditions, so that its flow is choked at every operating condition and does not respond to the acoustic pressure in the combustor; an injector that did respond would open a separate, simpler route to instability---injection coupling.
The injector is therefore acoustically stiff, and a mode that places a pressure antinode at the injector, as the \twoW\ of the long duct and the \oneW\ of the short one do, has little to act on.
A mode that places its velocity antinode there moves the flame directly.
Of the modes below $500$\,Hz in the two configurations only the \oneW\ of the long duct does so (\cref{tab:modes}): at the \oneW\ the acoustic pressure differs by a factor of $0.65$ between the two combustor sensors, $53$\,mm apart, whereas at the \twoW\ the two are equal to within $5\%$.

\begin{table}[tb]
\caption{Duct modes below 500\,Hz: dynamic pressure at the injector sensors relative to the duct sensor (negative: antiphase); half-power bandwidth $\Delta f$ of the duct-sensor spectrum and the net decay rate $\pi\Delta f$ it implies for a stable mode; decay rate $\alpha_e$ due to the exhaust from \cref{eq:alpha}; and in-phase response of the \chstar\ signal per unit combustor dynamic pressure near the injector, normalized by its largest value in each configuration.}
\label{tab:modes}
\centering
\begin{tabular}{llcccccc}
\toprule
Configuration & Mode & $f$, Hz & $p'_2/p'_1$ & $p'_3/p'_1$ & $\Delta f$, Hz ($\pi\Delta f$, s$^{-1}$) & $\alpha_e$, s$^{-1}$ & $\operatorname{Re}H_{qp}$ \\
\midrule
Unstable & $1W$ & $170$ & $+0.17$ & $+0.27$ & ---$^a$ & $0.3$--$1.1$ & $0.77$--$1.00$ \\
         & $2W$ & $264$ & $-1.23$ & $-1.30$ & $14.1$ ($44$) & $17$--$61$ & $0.13$ \\
         & $3W$ & $433$ & $+0.66$ & $+1.44$ & $5.1$ ($16$) & $1.6$--$5.7$ & $0.39$ \\
Stable   & $1W$ & $254$ & $-0.96$ & $-0.96$ & $4.8$ ($15$) & $21$--$61$ & $0.02$--$0.07$ \\
\bottomrule
\multicolumn{8}{l}{\footnotesize $^a$The instability; its line width does not measure a damping rate.}
\end{tabular}
\end{table}

The second is the acoustic loss.
The main loss of acoustic energy from the duct is through the exhaust, which removes energy in proportion to the square of the acoustic pressure at the holes (\cref{eq:W}), directly above the injector.
For the \oneW\ of the unstable configuration, with the exhaust near its pressure node, the decay rate $\alpha_e$ due to the exhaust is about $1$\,s$^{-1}$ (\cref{tab:modes}), whereas for the \twoW\ of the same configuration and for the \oneW\ of the stable one, with the exhaust near a pressure antinode, it is thirty to a hundred times larger, enough by itself to account for the measured damping of the \twoW.
The third mode has a small exhaust loss, like the first, and its bandwidth puts the other losses, at the walls, the pistons, and the injector, at about $10$\,s$^{-1}$; these grow with frequency, roughly as its square root for the viscous and thermal losses at the walls, so the value is an upper estimate for the \oneW, and the modes above the third, whose losses of this kind are still larger, are less likely to become unstable.
Of the three modes between $170$ and $270$\,Hz in the two configurations, the unstable \oneW\ is thus the only one that does not lose most of its energy through the exhaust.
The one-dimensional mode shapes reproduce the measured ratios of the acoustic pressure only roughly, so the estimates are reliable as to which modes have a node at the exhaust and not to better than a factor of two or three otherwise.

The third is the response of the flame.
The flame and the exhaust see the same acoustic pressure, so for every mode the ratio of the Rayleigh driving to the exhaust loss is proportional to $\operatorname{Re}H_{qp}$, the heat release in phase with the combustor dynamic pressure near the injector per unit of that pressure, whatever the mode shape; the last column of \cref{tab:modes} gives it, normalized within each record because the \chstar\ signal is uncalibrated.
It is about seven times larger at the unstable \oneW\ than at the \twoW, and small at the \oneW\ of the stable configuration, as coupling through the transverse velocity predicts: at the first the velocity moves the flame while the acoustic pressure near the injector is small, and at the others the acoustic pressure is large and the stiff injector does not respond to it.
The comparison also contains a dependence on frequency.
In the stable configuration the same ratio is large at $68$ and $137$\,Hz, where the duct has no mode but the flame and the acoustic pressure oscillate coherently for reasons we have not identified (the second is the EFD pair at $136$\,Hz), so the in-phase response of this flame is strong below about $200$\,Hz and weak above $250$\,Hz in both configurations.
An acoustically coupled instability needs, in addition, an acoustic mode at a frequency at which the flame responds.

The \oneW\ of the long duct is the only mode in either configuration that combines the three: a velocity antinode at the injector, an exhaust at a pressure node, and a frequency at which the flame responds strongly in phase.
The \oneW\ of the short duct, at nearly the same operating condition, has none of them, and that configuration is stable.

\subsection{Where the acoustic motion acts on the flow}
\label{sec:res-meanflow}

\begin{figure}[tb]
\centering
\includegraphics[width=\textwidth]{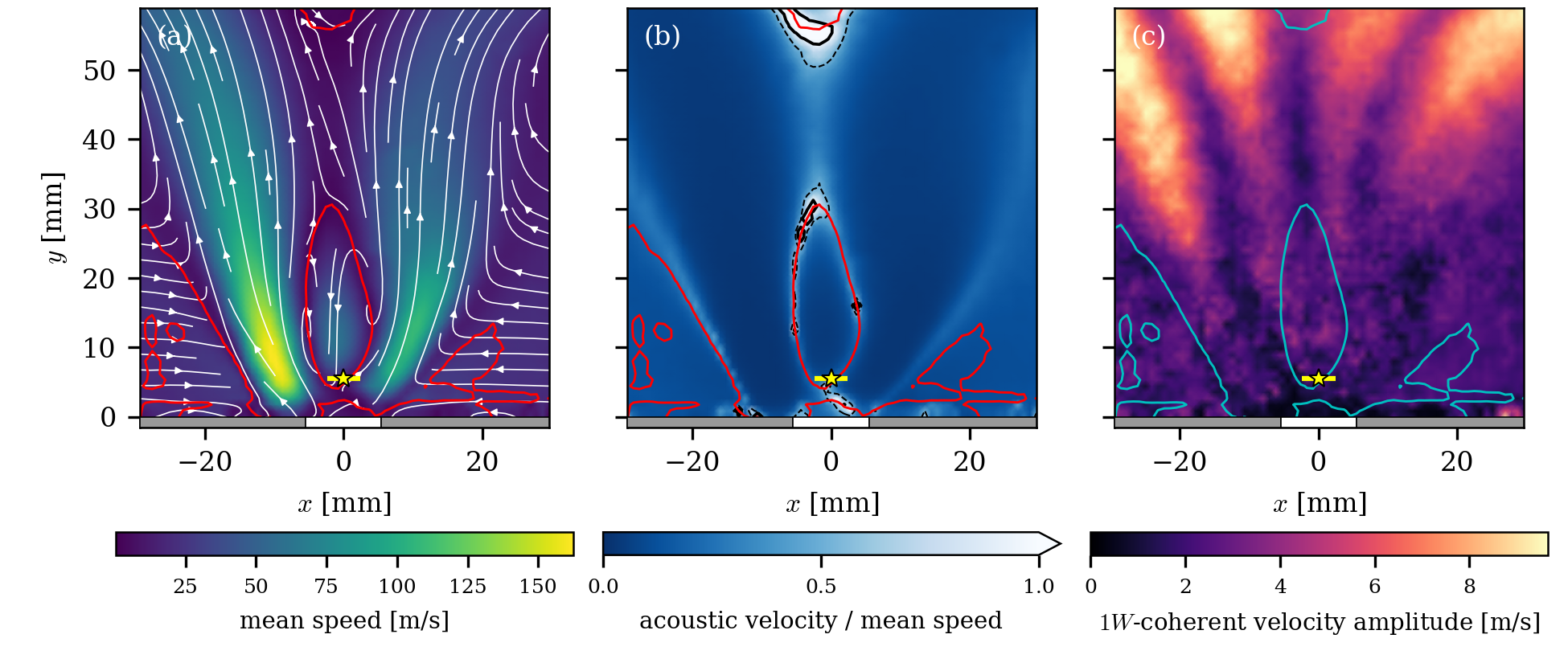}
\caption{Unstable configuration. a) Time-averaged speed and streamlines. b) Ratio of the acoustic velocity amplitude of the $1W$, 2.6\,m/s, to the local mean speed; the contours are at 0.5 (dashed) and 1 (solid). c) Amplitude of the velocity fluctuation at the $1W$, 150 to 190\,Hz, coherent with the combustor dynamic pressure near the injector. In each panel the closed contours are zero mean axial velocity, the star marks the upstream stagnation point of the inner recirculation zone, and the bar through it the acoustic displacement, $\pm2.4$\,mm.}
\label{fig:meanflow}
\end{figure}

\Cref{fig:meanflow}(a) shows the time-averaged velocity field of the unstable configuration.
The swirling jet leaves the nozzle as two branches in the measurement plane, with an inner recirculation zone between them that extends along the axis from $y = 5$ to $27$\,mm and is $9$\,mm wide, with a reverse flow of up to $48$\,m/s.
Its upstream stagnation point lies on the axis at the nozzle exit, and the flame is aerodynamically held there, in the low-speed flow below the recirculation zone (\cref{fig:cl}a).
The PIV is least reliable near the nozzle (\cref{sec:diagnostics}), and the velocities there should be read accordingly.
As is often the case for swirling flows, the jet is strongly asymmetric: the branch on the side away from the sensing section reaches $160$\,m/s near the nozzle and the other $90$ to $110$\,m/s.

The acoustic velocity can perturb the flow where it is comparable with the mean velocity, and \cref{fig:meanflow}(b) maps the ratio of the two.
In the jets the acoustic velocity is a few percent of the mean velocity; it rises to order one only where the mean flow stagnates, and the spot that matters for the mechanism is the upstream stagnation point of the recirculation zone, where the flame is held and the shear layers start.
The ratio exceeds one half only in the slow flow between the nozzle exit and that point, around which the mean velocity rises to the speed of the jets within a few millimeters, so the reach of the acoustic motion is better measured by its displacement.
Each cycle of the \oneW\ sweeps the stagnation point that anchors the flame sideways by about $2$\,mm, $0.36$ to $0.44$ nozzle radii in each direction, a first-order perturbation at the anchor point and close to the largest displacement that the mode produces anywhere in the duct, since its velocity antinode is near the injector.

\subsection{Response of the heat release}
\label{sec:res-cl}

\begin{figure}[tb]
\centering
\includegraphics[width=\textwidth]{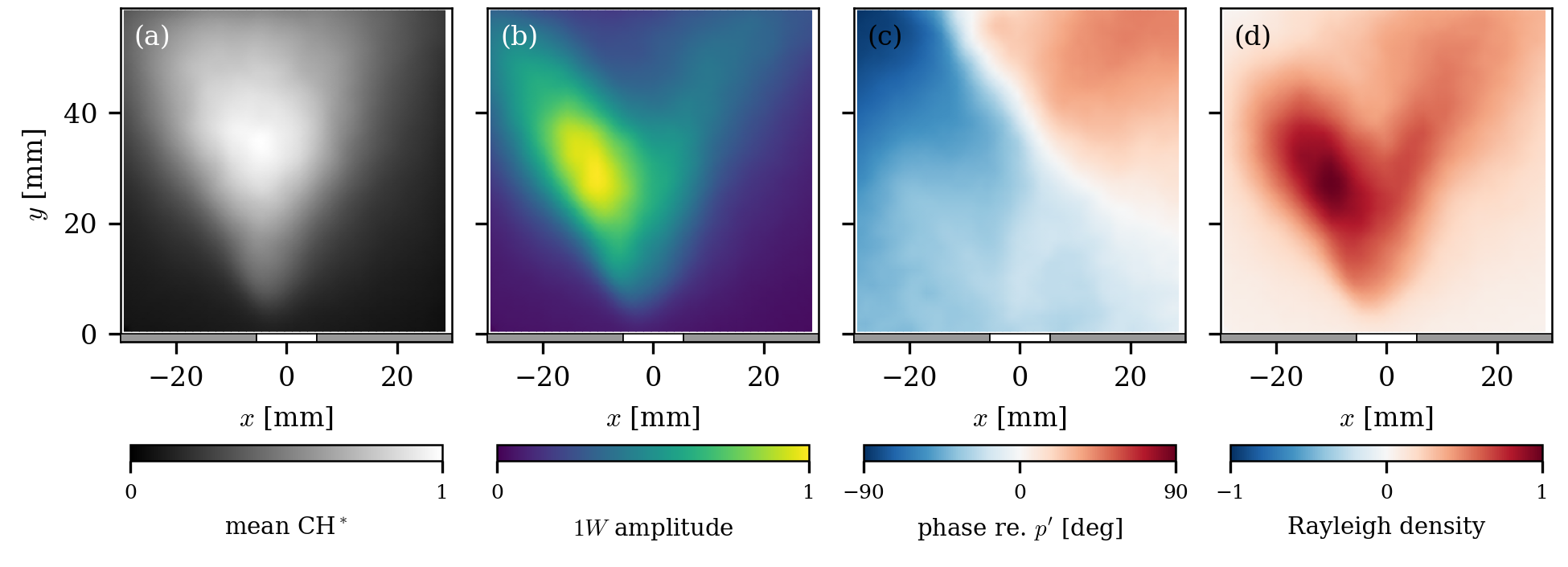}
\caption{\chstar\ chemiluminescence in the unstable configuration. a) Time average. b) Amplitude of the fluctuation between 150 and 190\,Hz. c) Phase by which the fluctuation leads the combustor dynamic pressure near the injector. d) Local Rayleigh density. Panels a, b, and d are scaled by their maxima.}
\label{fig:cl}
\end{figure}

\Cref{fig:cl} maps the \chstar\ signal of the unstable configuration at the \oneW.
Its sum over the image is coherent with the combustor dynamic pressure near the injector, and so is every pixel: the whole flame takes part.

The heat release lags the combustor dynamic pressure near the injector by about $25^\circ$, the phase at every pixel is within $90^\circ$ of the acoustic pressure (\cref{fig:cl}c), and the local Rayleigh density is positive everywhere (\cref{fig:cl}d): the normalized Rayleigh index of \cref{eq:rayleigh} is about $+0.7$, and the heat release drives the mode at nearly all locations, as found in \cite{Quinlan-2022}.
With the corrected time alignment the index is positive at the other duct modes of both configurations as well, contrary to the negative value reported for the $258$\,Hz mode in \cite{Quinlan-2022} before the alignment was corrected, so its sign is necessary for the instability but does not select the unstable mode; what selects it is the strength of the driving relative to the losses (\cref{sec:res-why}).

The response is one-sided.
The time-averaged flame (\cref{fig:cl}a) is a V anchored at the nozzle and tilted toward the faster branch of the jet, as swirl flames commonly are, and the fluctuation at the \oneW\ (\cref{fig:cl}b) is concentrated on that branch, which carries nearly two-thirds of the fluctuation power.
A symmetric flapping about the axis would put the two sides in antiphase; instead they differ in phase by about $70^\circ$, the dominant side lagging the acoustic pressure by about $50^\circ$ and the other leading it slightly, as the in-phase and quadrature components of the mode of \cref{fig:mode170} also show.
The response is thus a flapping carried mainly by one branch of an asymmetric flame, near the maximum of the acoustic pressure, with a phase that progresses across the flame.
A line-of-sight view cannot by itself distinguish a one-sided response from a disturbance that travels around the axis, but the planar OH signal agrees with the \chstar\ signal (\cref{fig:mode170}), so the response can be read as a local perturbation of the heat release.
In the quadrature components of the mode the transverse velocity has one sign over the field, so it is directed toward the axis of the flame on one branch of the V and away from it on the other, and the heat release is below its mean on the first branch and above it on the second: the acoustic velocity pushes one branch of the flame inward and carries the other outward \cite{Quinlan-2022}.

The timing of the response is that of a disturbance convected from the anchoring point.
The summed heat release lags the acoustic velocity by about a third of a period, $1.9$\,ms, and at the convection speed of the structures of the shear layer, about $36$\,m/s \cite{Quinlan-2015}, a disturbance created at the anchoring point travels $68$\,mm in that time: through the whole of the flame, which extends beyond the field of view, and most of the way to the exhaust holes, $74$ to $82$\,mm away.
The phase at a single frequency does not exclude a response proportional to the acoustic displacement with little delay, which would also fall at the maximum of the acoustic pressure.

\subsection{The shear-layer mode near 1 kHz}
\label{sec:res-khz}

The hydrodynamic instability expected to amplify the acoustic perturbation is that of the injector shear layer.
In the DMD of the velocity, OH, and \chstar\ data of the unstable configuration it appears as the leading mode, at $984$\,Hz, a pair of helical vortices spinning around the inner recirculation zone \cite{Quinlan-2017}.
The flame lies almost on the shear layer, so the mode modulates the heat release as well as the velocity, as its OH and \chstar\ parts show.
It is not a motion at a single frequency: the vortices grow as they are carried downstream and their frequency falls, so the structure is represented by a succession of modes at neighboring frequencies, each strongest a little farther downstream \cite{Quinlan-2015}.
The mode is therefore narrowband and not strictly periodic.
Its period is the analysis period of \cref{sec:res-efd}.
Three observations characterize it.

\begin{figure}[tb]
\centering
\includegraphics[width=\textwidth]{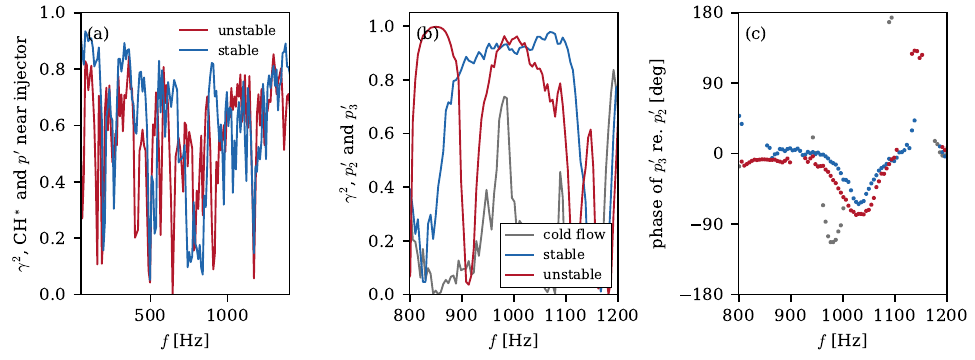}
\caption{The band near 1\,kHz. a) Coherence between the \chstar\ signal summed over the image and the combustor dynamic pressure near the injector. b) Coherence between the two injector sensors. c) Phase of $p'_3$ relative to $p'_2$ where their coherence exceeds 0.3.}
\label{fig:khz}
\end{figure}

First, the mode is present in cold flow.
The standoff tubes modify the peak of the dynamic pressure near $1$\,kHz at the injector sensors (\cref{sec:spectra}), but in cold flow the two injector sensors, on opposite sides of the nozzle, record far more than the duct sensor in that band and are only weakly coherent with each other, with a large phase difference (\cref{fig:khz}b and c), whereas two nearly identical tubes listening to a common sound field would be almost perfectly coherent and nearly in phase.
What they hear in cold flow therefore consists mostly of pressure fluctuations generated separately near each mouth, which points to a mode of the shear layer of the swirling jet that exists without combustion.

Second, with combustion the heat release takes part: the injector sensors then become coherent with each other and with the duct sensor, the tubes amplifying a sound field, and the \chstar\ signal summed over the image is coherent with the combustor dynamic pressure near the injector across the band (\cref{fig:khz}a), in both configurations.

Third, the mode is not phase-locked: synchronous averaging of the velocity and scalar fields finds no component that repeats with a period near $1$\,ms over the record, and a DMD of the velocity alone represents it only as a strongly damped mode, as it does a narrowband motion.

\subsection{From the acoustic perturbation to the heat release}
\label{sec:res-coupling}

The perturbation of the anchoring point by the \oneW\ can reach the heat release along the shear layer in two ways.

The first is a direct response at $170$\,Hz.
The part of the local velocity that is coherent with the acoustic pressure at the \oneW\ (\cref{fig:meanflow}c) is at the acoustic level near the nozzle and grows along both branches of the jet, by a factor of two to three over the field, to its largest values toward the exhaust: the flow responds directly, by a sideways flapping of the jets that grows downstream, but weakly.
At $170$\,Hz the Strouhal number based on the nozzle diameter and the jet velocity is more than an order of magnitude below the preferred frequencies of jets and their shear layers \cite{Ho-1984}, so the shear layer is not expected to amplify a disturbance at this frequency strongly, and there is no indication that it does.

The second is a modulation of the shear-layer mode.
If the displacement of the anchoring point by the \oneW\ modulates the strength of the activity near $1$\,kHz, the heat release that this activity carries is modulated at $170$\,Hz without any amplification at $170$\,Hz.
Such a modulation adds components at $f \pm 170$\,Hz whose phases are tied to those of the \oneW\ and of the component at $f$, a tie that the bicoherence detects pair by pair and the envelope test summed over the band (\cref{sec:spectral}).

In the dynamic pressure the coupling is clear.
With the \oneW\ taken from the duct sensor and the other two components from one injector sensor, $b^2$ exceeds its $95\%$ level at a quarter to a third of the frequencies between $500$ and $1500$\,Hz, several times the number expected by chance and far more than along lines at frequencies where there is no mode.
Among the coupled components are the sixth duct mode and the activity near $1$\,kHz, but not the harmonics of the \oneW.
The coupling is absent in the stable configuration, which has no mode at $170$\,Hz, stronger in the second half of the record, where the \oneW\ is nearly twice as strong, and present at the duct sensor, which has no standoff tube.

In the velocity, whose record is a fifth as long, only the envelope test is used.
Of $110$ combinations of position and component over the image, nine exceed the $95\%$ level, where five or six would by chance, all at the downstream edge of the field, on both branches of the jet; at the stagnation point at the nozzle the envelope is not coherent with the \oneW.

The modulation is therefore supported clearly by the dynamic pressure, which does not locate it, and weakly by the velocity, near the top of the field, where the flapping of the jet, which carries the shear layer back and forth across a fixed point, may also contribute.
The bicoherence is symmetric in its components and does not say whether the \oneW\ modulates the faster activity or is modulated by it; the second reading is taken up in \cref{sec:res-param}.

Both responses are strongest where the field of view ends, $17$\,mm below the exhaust holes.
The disturbances carried along the shear layers reach the holes after a delay of $2.0$ to $2.3$\,ms at $36$\,m/s, a little more than a third of a period of the \oneW, and their interaction with the holes and the upper wall is a further source of sound, through the unsteady heat release of the enhanced mixing there and through the acceleration of vorticity into the holes \cite{Quinlan-2022,Poinsot-1987}.
That region is outside the measurements, and its contribution to the feedback is undetermined, but seems likely to contribute to this instability.

\subsection{Tests for a parametric coupling}
\label{sec:res-param}

A parametric coupling between the \oneW\ and the shear-layer mode would leave two kinds of trace: content in the Floquet mode at $170 \mp 1000$\,Hz, locked to the period of the faster motion, and a fixed relation between the phases of the \oneW, the faster motion, and the components at the sums and differences of their frequencies.
The harmonics of the EFD mode (\cref{eq:harmonics}) test the first, with the time derivatives of the dynamic pressure band-passed to $1000$\,Hz so that the mode can hold the duct modes up to the sixth.
With the record resampled so that the sixth, the fifth, or the fourth duct mode falls on the first harmonic of the \oneW, the harmonic absorbs that mode, the sixth somewhat more than the others but within the spread of the controls, so the duct modes enter largely by aliasing, and the harmonics neither establish nor exclude a combination resonance between the \oneW\ and the sixth duct mode pumped near $1$\,kHz \cite{Nayfeh-1979}.
Nor do the two modes wander in opposite senses, as a pump of steady frequency would make them.

The bicoherence, which uses the phases themselves, shows the coupling that these tests do not see as a lock.
The dynamic pressure spectrum has narrow lines at the sum and the difference of the frequencies of the \oneW\ and the sixth duct mode, much stronger than the nonlinearity of the sound field produces at the amplitude of the \oneW, and the phase of the component near $1020$\,Hz follows the sum of the phases of the two modes as they wander; the triad is one of many coupled pairs along the line of the \oneW.
The \oneW\ and the activity near $1$\,kHz are therefore coupled, more strongly than the acoustics alone can explain, and the coupling is spread over the band rather than locked to one period or one triad, which is the form that a modulation by the shear-layer mode, a succession of slowly changing modes rather than one steady oscillation, is expected to take (\cref{sec:disc-parametric}).

%% file: sections/05_discussion.tex
\section{Discussion}
\label{sec:discussion}

\subsection{The proposed mechanism}
\label{sec:disc-mechanism}

\begin{figure}[tb]
\centering
\begin{tikzpicture}[font=\small,>=Stealth,
  box/.style={draw,rounded corners=2pt,align=center,inner sep=4pt,text width=3.7cm,minimum height=1.5cm}]
  \node[box] (A) at (0,0)      {First transverse mode, 170\,Hz: velocity antinode at a stiff injector};
  \node[box] (B) at (5.4,0)    {Flame anchor at the stagnation point displaced by about 0.4 nozzle radii};
  \node[box] (C) at (10.8,0)   {Shear layer carries and amplifies the disturbance: directly at 170\,Hz, or by modulation of its mode near 1\,kHz};
  \node[box] (D) at (10.8,-3.1) {Heat release responds over the flame, mainly on one branch, near the acoustic pressure maximum};
  \node[box] (E) at (5.4,-3.1)  {Disturbances reach the exhaust holes: further sound (outside the field of view)};
  \draw[->,thick] (A) -- node[above]{acoustic} node[below]{velocity} (B);
  \draw[->,thick] (B) -- (C);
  \draw[->,thick] (C) -- node[left,align=right]{delay of about\\a third of a period} (D);
  \draw[->,thick] (D) -- node[above]{carried on} (E);
  \draw[->,thick,dashed] (E.west) -| node[pos=0.25,above]{sound} ([xshift=0.9cm]A.south);
  \draw[->,thick] (D.south) -- ++(0,-0.6) -| node[pos=0.25,below]{Rayleigh driving} ([xshift=-0.9cm]A.south);
  \draw[->,thick] (A.north) -- ++(0,0.7) node[above,align=center]{small acoustic loss:\\exhaust at the pressure node};
\end{tikzpicture}
\caption{The proposed feedback loop of the instability. The dashed link is not measured.}
\label{fig:loop}
\end{figure}
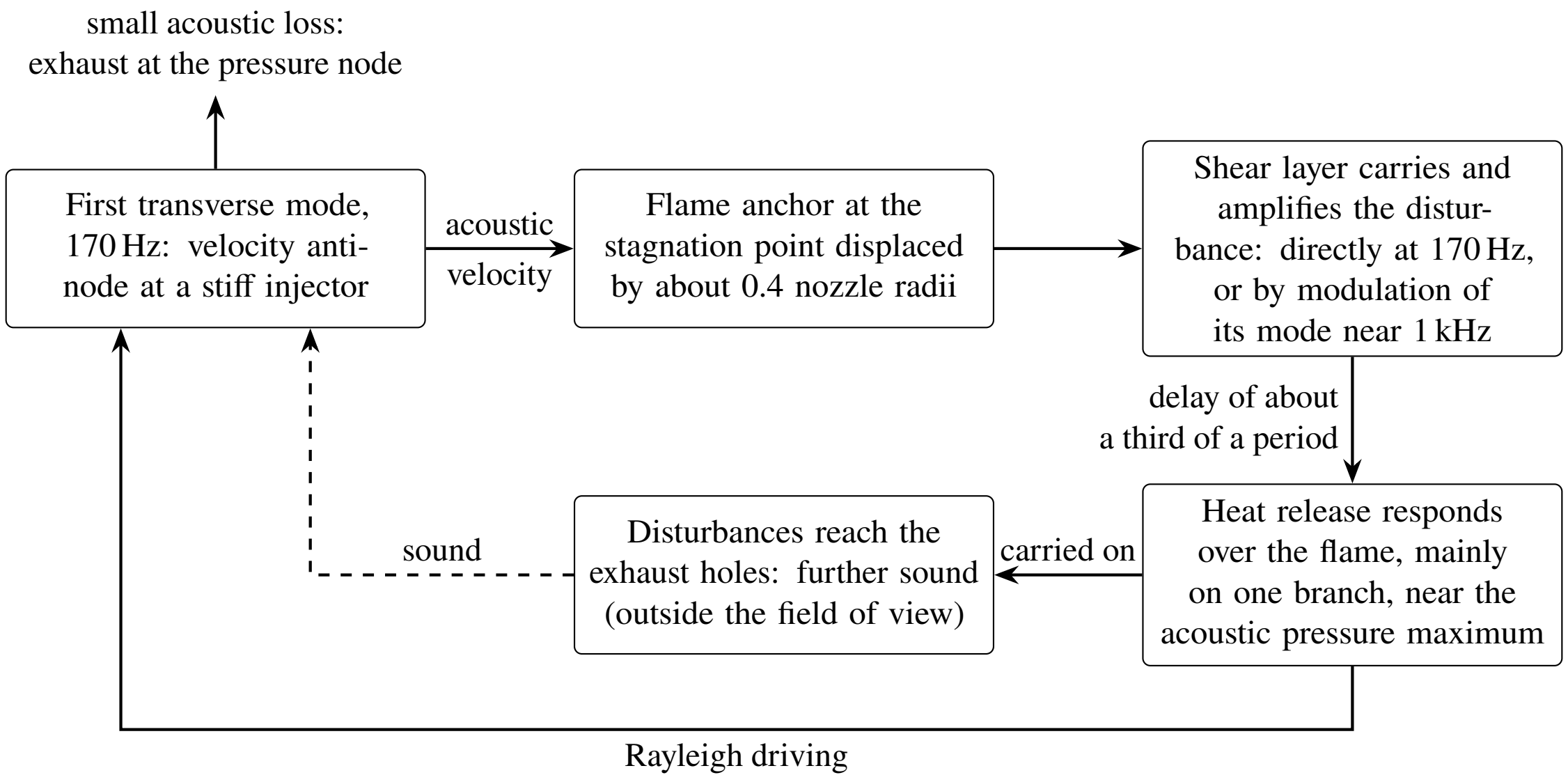

The results assemble into the feedback loop of \cref{fig:loop}.
We take its links in turn, with the evidence for each.

\emph{The mode and its coupling.}
The instability is the first transverse mode of the long duct, whose velocity antinode lies at the acoustically stiff injector, so that the mode acts on the flame through its transverse velocity, which the PIV resolves (\cref{sec:res-mode,sec:res-why}).

\emph{Where the perturbation enters.}
That velocity is comparable with the mean flow only where the flow stagnates, and at the stagnation point where the flame is held it displaces the anchor by about $0.4$ nozzle radii in each direction (\cref{sec:res-meanflow}).

\emph{Amplification in the shear layer.}
The perturbation is carried along the shear layer, whose mode near $1$\,kHz modulates the heat release (\cref{sec:res-khz}), either by a direct flapping at $170$\,Hz, which is observed but grows little, or by a modulation of the activity near $1$\,kHz, which the dynamic pressure supports clearly (\cref{sec:res-coupling}).

\emph{The heat release and its timing.}
The heat release responds over the whole flame near the maximum of the acoustic pressure, about a third of a period after the acoustic velocity, the time a disturbance takes to cross the flame, and returns energy to the mode (\cref{sec:res-cl}).

\emph{The exhaust.}
The disturbances go on to the exhaust holes, where their interaction with the holes and the wall is a further source of sound, by unsteady heat release in the enhanced mixing there and by the acceleration of vorticity into the holes \cite{Quinlan-2022}; that region lies beyond the field of view, and this link is not measured.

\emph{The losses.}
The exhaust, the main sink of acoustic energy in the duct, lies at the pressure node of the \oneW, which therefore loses thirty to a hundred times less energy through it than the modes with a pressure antinode there (\cref{sec:res-why}).

\subsection{Whether the instability is parametric}
\label{sec:disc-parametric}

The shear-layer mode near $1$\,kHz is a pair of helical vortices rotating about the inner recirculation zone (\cref{sec:res-khz}) \cite{Quinlan-2017}, whereas the \oneW\ is fixed in the frame of the duct.
A structure that rotates through a fixed acoustic mode presents it with a medium that varies periodically: the position of the flame, the local velocity and sound speed, and the gain with which the heat release responds, all of which enter the equations of the acoustic mode as coefficients.
Seen from the acoustic mode, the system then has periodic coefficients, the setting of Floquet theory and of the pendulum of \cref{eq:mathieu}, and the instability is parametric if that periodic variation is what makes the mode grow.
The same rotation is the reason why measurements in a fixed plane must be gathered over a whole cycle to describe the flow: one image shows one section of the structure, and the images over one revolution, about ten of them, show all of it (\cref{sec:efd}).

Such an instability would differ from the parametric instabilities established in combustion, in which the acoustic field, imposed or self-excited, modulates the flame front \cite{Markstein-1955,Searby-1991} or the precessing vortex core \cite{Terhaar-2016} and the response is subharmonic: here the periodic motion would be hydrodynamic, the mode that grows acoustic, and the ratio of their frequencies about six.
In a self-excited system, however, the distinction between a pump and the oscillator that it excites is one of description rather than of physics: the coupled oscillators of the companion paper \cite{Quinlan-2026-FD}, an autonomous analog of Mathieu's equation, exchange energy in both directions, and an instability of their joint periodic motion is parametric without either oscillator being a pump.
The difference-frequency component observed in swirl flames by Steinberg et al.\ \cite{Steinberg-2010} and Moeck et al.\ \cite{Moeck-2012} belongs to the same family, a nonlinear interaction of two oscillations and a parametric modulation of one by the other being two descriptions of one coupling.
A single oscillator modulated at six times its natural frequency is far from every tongue of \cref{eq:mathieu}; at such a ratio the theory of parametric excitation offers a combination resonance with a second mode whose frequency makes up the difference, or a modulation without resonance that changes the effective damping at second order \cite{Nayfeh-1979}.

Four things make the question hard to decide from these data.
The loop may close without a parametric step: if the shear layer responds directly at $170$\,Hz, or if the activity near $1$\,kHz is only a carrier whose modulation at $170$\,Hz conveys the disturbance to the heat release, the feedback is closed at the acoustic frequency, as in the vortex-driven instabilities described by Poinsot et al.\ \cite{Poinsot-1987} and by Schadow and Gutmark \cite{Schadow-1992}.
The bicoherence is symmetric in its components, so that a modulation of the activity near $1$\,kHz by the \oneW, which closes the loop at the acoustic frequency, and a modulation of the \oneW\ by that activity, which makes the instability parametric, leave the same trace.
The shear-layer mode is narrowband and not strictly periodic, a succession of slowly changing modes, so that a modulation by it is spread over its band, and the tests of \cref{sec:res-param}, which look for a lock to one period or to one triad, cannot isolate it.
And the identification of the mode is not by itself evidence of periodic dynamics: EFD returns the modes of a time-invariant system as well, the mode is the same whatever analysis period is assumed, and Hankel-DMD finds it too (\cref{sec:res-efd}).

Against these, three observations point to a parametric instability.
First, the two frequencies are coupled, and not through the acoustics alone: in the dynamic pressure the \oneW\ is phase-coupled to much of the activity between $600$ and $1500$\,Hz, not through its harmonics, more strongly when its amplitude is larger, and not at all in the stable configuration, and the lines at the sum and difference of the frequencies of the \oneW\ and the sixth duct mode are far stronger than the nonlinearity of the sound field produces (\cref{sec:res-coupling,sec:res-param}).
Second, in the composite data, which contain the flow and the flame as well as the dynamic pressure, the instability is separated only by methods that see a whole cycle of the rotating structure.
DMD and FD, which see one instant, miss it at the rank of the analysis, although the \oneW\ holds nearly half of the variance of the dynamic pressure at the duct sensor, and the dynamic pressure and its rate of change at three points can determine the state of the lowest acoustic modes at one instant; FD finds it only at twice the rank or more, and DMD and FD find it only when the band of the dynamic pressure is narrowed to below about $400$\,Hz, to little more than the \oneW\ and \twoW, which in effect tells them where to look (\cref{sec:res-efd}).
EFD finds it, and Hankel-DMD finds it only with a window of at least one cycle of the $1$\,kHz mode, below which it returns a mixture of the two modes (\cref{sec:efd}).
What must be known to separate and predict the \oneW\ in these data is thus the state of the rotating structure, which the measurements determine only over one of its cycles, as expected if the \oneW\ evolves together with that structure as a Floquet mode of a system whose coefficients vary with it.
Third, the instability is selective.
The stiff injector closes the injection-coupled route, and of the hundreds of piston configurations tried, many with the injector near a velocity antinode and so the exhaust near a pressure node, only this one was unstable (\cref{sec:rig}).
Coupling through the transverse velocity at an acoustically stiff injector and a small loss at the exhaust are therefore not sufficient; something selects $170$\,Hz in this duct.
A resonance between the acoustic modes and the shear-layer mode would do so, for instance the combination of the \oneW\ and the sixth duct mode, whose frequencies add to $1020$\,Hz, within the band of the shear-layer mode, and whose triad with that band is coupled; what selects this configuration has yet to be established.

We therefore suggest that the instability is a parametric, acoustically coupled combustion instability, in which a hydrodynamic mode of the injector shear layer, rotating through a fixed acoustic mode, supplies the periodic variation of the coefficients that makes the acoustic mode grow.
The data make this the most consistent reading of what was measured, but they do not prove it: that requires the modulation to be shown to change the energy balance of the acoustic mode.

%% file: sections/06_conclusions.tex
\section{Conclusions}
\label{sec:conclusions}

Simultaneous measurements of velocity, OH fluorescence, \chstar\ chemiluminescence, and dynamic pressure in a swirl-stabilized combustor with an adjustable transverse duct were analyzed with the ensemble Floquet decomposition.
EFD identified the instability at $170$\,Hz, which DMD and the single-phase Floquet decomposition of the same snapshots did not, as the first transverse mode of the duct, and returned it as one mode in which the four measurements share one multiplier.
With that mode the instability was traced to a feedback loop.
The transverse acoustic velocity at an acoustically stiff injector displaces the stagnation point that anchors the flame by about $0.4$ nozzle radii, the disturbance is carried along the shear layer of the swirling jet, and the heat release responds over the whole flame, mainly on one branch, near the maximum of the acoustic pressure; the exhaust, at the pressure node of the mode, removes thirty to a hundred times less energy from it than from its neighbors.
The mode is phase-coupled to the mode of the injector shear layer near $1$\,kHz, a pair of helical vortices that rotates through the fixed acoustic mode.

These results change what is known about this instability.
The conditions that an acoustic analysis would name as its cause, coupling through the velocity at a stiff injector and a small loss at the exhaust, were met in many other configurations of the same rig that were stable, so they are necessary but not sufficient.
What singles out the unstable configuration is a coupling between the acoustic mode and a hydrodynamic mode six times faster, which rotates through it and so presents it with periodically varying coefficients, and the acoustic mode can be separated in these data only together with the state of that rotating structure.
We suggest that the instability is therefore a parametric, acoustically coupled combustion instability, of a kind different from those established in combustion: the periodic motion is a hydrodynamic mode of the injector flow, the mode that grows is an acoustic mode of the chamber, and the two are coupled at a ratio of about six rather than two.
The data support this reading without proving it; the loop may still close at the acoustic frequency alone.

If instabilities of this kind exist, they matter beyond this rig.
Combustors of rocket engines and gas turbines are made stable largely by closing the known routes to instability: stiff injectors against injection coupling, damping tuned to the acoustic modes, and time lags kept away from the acoustic frequency.
As those routes are closed, instabilities whose feedback passes through a hydrodynamic mode of the injector flow at a frequency far from the acoustic one may be among the next to appear, and the configuration studied here, stable in every other setting, shows how narrow the conditions for them can be.
They would also respond differently to remedies.
A change that adds acoustic damping or shifts a time lag treats the acoustic part of the loop and may leave the parametric part, the modulation by the hydrodynamic mode, in place, whereas a change to that mode, to its frequency, its strength, or its rotation, may remove the instability without any change to the acoustics.

The ensemble Floquet decomposition is what made the analysis possible.
From $0.85$\,s of velocity, OH, and \chstar\ images and three dynamic pressure signals, about $1.4\times10^{8}$ numbers in which DMD of the same snapshots found nothing at $170$\,Hz, it extracted the instability as one mode in which the four measurements are consistently related, and Hankel-DMD did the same once its window was chosen from the period.
The reason is general.
EFD requires that the measurements made during one period determine the state, not those made at one instant, which is the situation of planar and line-of-sight measurements of three-dimensional flows, and it applies whether the periodicity lies in the dynamics or in the observation, as when a rotating structure is seen in a fixed plane.
It may therefore serve as a tool for systems with periodic coefficients and for periodically observed systems generally, in which what a single snapshot cannot show becomes visible over one cycle.
Three cautions apply to its use.
A harmonic of an EFD mode also absorbs an unrelated oscillation at its alias frequency, so its content is evidence of a coupling only if it is absent at control periods; EFD recovers the modes of a time-invariant system as well, so that it finds a mode is not by itself evidence that the dynamics are periodic; and with noisy data its moduli measure the coherence of a mode, not its growth.
The identification also depends on choices, the weight and the band of the dynamic pressure signals above all, which amount to telling the method where to look; this work and the companion paper \cite{Quinlan-2026-FD} show how these choices can be made.